%% file: main.tex
\documentclass[%
 reprint,
 superscriptaddress,
 altaffilletter,
 amsmath,amssymb,
 aps,
floatfix,
]{revtex4-2}

\usepackage{graphicx}% Include figure files
\usepackage{dcolumn}% Align table columns on decimal point
\usepackage{bm}% bold math
\usepackage{hyperref}% add hypertext capabilities
\usepackage[mathlines]{lineno}% Enable numbering of text and display math
\usepackage{mathtools}
\newcommand{\nueXeCC}{\ensuremath{\nu_e}CC-Xe}
\newcommand{\XeTCC}{\ensuremath{\nu_e-{}^{132}}XeCC}
\newcommand{\XeFCC}{\ensuremath{\nu_e-{}^{134}}XeCC}
\newcommand{\XeSCC}{\ensuremath{\nu_e-{}^{136}}XeCC}
\newcommand{\veES}{\ensuremath{\nu_e-e^-}}
\newcommand{\nuES}{\ensuremath{\nu e ES}}

\begin{document}
\renewcommand{\arraystretch}{1.2}

\preprint{APS/123-QED}

\title{Supernova electron-neutrino interactions with xenon in the nEXO detector}

\date{\today}

\input{authors}

\begin{abstract}
Electron-neutrino charged-current interactions with xenon nuclei were modeled in the nEXO neutrinoless double-$\beta$ decay detector ($\sim$5 metric ton, 90\% ${}^{136}$Xe, 10\% ${}^{134}$Xe) to evaluate its sensitivity to supernova neutrinos. Predictions for event rates and detectable signatures were modeled using the MARLEY (Model of Argon Reaction Low Energy Yields) event generator. We find good agreement between MARLEY's predictions and existing theoretical calculations of the inclusive cross sections at supernova neutrino energies. The interactions modeled by MARLEY were simulated within the nEXO simulation framework and were run through an example reconstruction algorithm to determine the detector's efficiency for reconstructing these events. The simulated data, incorporating the detector response, were used to study the ability of nEXO to reconstruct the incident electron-neutrino spectrum and these results were extended to a larger xenon detector of the same isotope enrichment. We estimate that nEXO will be able to observe electron-neutrino interactions with xenon from supernovae as far as 5-8\,kpc from Earth, while the ability to reconstruct incident electron-neutrino spectrum parameters from observed interactions in nEXO is limited to closer supernovae. 
\end{abstract}

\maketitle
%\tableofcontents

\section{Introduction}
%Supernovae physics with neutrinos
The predicted rate of core-collapse supernovae (CCSNe) within our Galaxy is estimated to be a few per century, with large uncertainties~\cite{Cappellaro1993,Tammann1994,Adams2013,Agafonova2015,Rozwadowska2021}. When a CCSN occurs, $\sim$99\% of its gravitational binding energy is emitted in the form of neutrinos~\cite{Mirizzi2016}, which can both provide an early alert to astronomers that a supernova has occurred~\cite{AlKharusi2021}, as well as provide valuable information about the explosion dynamics. Neutrinos and antineutrinos of all flavors are expected to be produced during a core-collapse supernova, with the energy and flavor content varying throughout the explosion. In particular, electron neutrinos are expected to be produced in abundance during the infall phase and neutronization burst of a CCSN~\cite{Janka2017}. By studying the energy spectra of the different flavors of neutrinos produced during a CCSN, it is possible to image the interior dynamics of the collapsing star, as different flavors will decouple from thermal equilibrium at different depths. Studying the time, energy, and flavor composition of detected neutrinos can provide insight into the neutrino mass ordering~\cite{Dighe2008}, set bounds on the neutrino mass~\cite{Beacom1998}, and test nonstandard neutrino interactions and physics beyond the Standard Model~\cite{Schramm1990,Raffelt1999}.

Different detection channels are well suited for measuring the different flavors of supernova neutrinos. Inverse-$\beta$ decay on hydrogen (IBD) is commonly used for detection of electron antineutrinos, coherent elastic neutrino-nucleus scattering (CE$\nu$NS) for detection of the neutral-current (NC) component, and neutrino-electron elastic scattering or charged-current (CC) scattering with nuclei for the electron-neutrino component. Reference~\cite{AlKharusi2021} reviews detection channels and detector capabilities.

Two examples of detectors with electron-neutrino charged-current ($\nu_e$CC) sensitivity are HALO~\cite{Duba2008} and the future DUNE experiment~\cite{Abi2021}, which will utilize $\nu_e$CC events on lead and argon, respectively, to search for supernovae. As $\nu_e$CC cross sections tend to scale with increasing neutron excess in the interacting nucleus~\cite{Mintz2001}, smaller detectors with neutron-rich targets such as xenon can still have sensitivity to supernova electron neutrinos. Additionally, future detectors have been proposed for neutrinoless double-$\beta$ decay ($0\nu\beta\beta$) and weakly interacting massive particle dark-matter searches with masses ranging from 30 metric tons to 3 metric kt~\cite{Aalbers2016,Avasthi2021,Wang2023}, which would have increased sensitivity.

The nEXO detector will search for neutrinoless double-$\beta$ decay using a single-phase liquid xenon time-projection chamber (TPC) enriched to 90\% in the ${}^{136}$Xe isotope~\cite{Adhikari2022}. nEXO’s sensitivity to $0\nu\beta\beta$ decay is robust against backgrounds and ``unknown unknowns” due to the use of multiple observables for signal-to-background discrimination. This includes, but is not limited to, its design energy resolution of $<$1\% at the large ${}^{136}$Xe $Q$ value of 2.458 MeV. These factors result in an estimated $0\nu\beta\beta$ half-life sensitivity of $1.35\times10^{28}$\,yr at 90\% confidence level after 10\,yr of data taking. Many of the properties that make nEXO ideal for detecting $0\nu\beta\beta$ also give it sensitivity to supernova electron-neutrino CC interactions with xenon (\nueXeCC{}) such as its large homogeneous xenon volume, location deep underground, low-background design, optimization for MeV-scale physics, and planned 10-yr exposure. Existing calculations~\cite{Divari2013,Pirinen2019,Bhattacharjee2022} of \nueXeCC{} interactions typically focus on interaction rates, but not on how many of those interactions would be detectable. To realistically study these interactions in nEXO, particles produced by electron-neutrino interaction must be modeled along with the detector response to their interactions.

Section~\ref{Sect:nEXO} describes the nEXO detector and relevant details for detecting supernovae. Section ~\ref{Sect:MARLEY} focuses on the modifications of the MARLEY (Model of Argon Reaction Low Energy Yields) event generator~\cite{Gardiner2021} required for modeling \nueXeCC{} interactions; here MARLEY's cross section predictions are compared with existing theoretical calculations. Section~\ref{Sect:reconstruction} details the simulation of events predicted by MARLEY and the reconstruction algorithm employed to predict the visible spectrum nEXO will observe as a result of electron-neutrino interactions. Finally, Sec.~\ref{Sect:parameterization} studies the ability of the nEXO detector to reconstruct the incident supernova electron-neutrino parameters. These results are extended to a larger 300 metric ton detector of similar isotopic enrichment. 

%%%%%%%%%%%%%%%%%%%%%%%%%%%%%%%%%%%%%%%%%%%%%%%%%%%%%%%%%%%%%%%%%%%%%%%%%%%%%%%%%%%%%%%%%%%%%
\section{The nEXO detector}\label{Sect:nEXO}
The conceptual design for the nEXO detector features $\sim$5 metric tons of xenon enriched to 90\% in the isotope ${}^{136}$Xe. The remaining 10\% consists mainly of ${}^{134}$Xe~\cite{Adhikari2022}. The liquid xenon will be contained within a cylindrical copper vessel with an inner diameter of 127.7\,cm and height of 127.7\,cm. Additional details on the detector can be found in Ref.~\cite{Adhikari2022}. The collaboration is planning to locate the detector deep underground at the SNOLAB Cryopit, with $\sim$6,000-m water-equivalent overburden, where the cosmic muon rate is 0.27 muons $\mbox{m}^{-2}\mbox{ day}^{-1}$~\cite{Jillings2016}. The detector will be surrounded by a 12.3-m-diameter 13.3-m-high water tank, acting as an active muon veto as well as providing passive shielding from external backgrounds. In addition to being sensitive to supernova electron-neutrino interactions in the xenon target, the detector will also potentially be sensitive to supernova electron-antineutrino IBD interactions in the water tank~\cite{AlKharusi2021}.

Particle interactions in the xenon produce vacuum ultraviolet scintillation light, which will be detected by silicon photomultipliers (SiPMs) lining the vertical walls of the detector, with the number of optical photons (OPs) serving as the measure of light intensity. Simultaneously, particle interactions will ionize xenon atoms, producing electrons, which are subsequently drifted to the detector anode with an applied electric field of 400\,V/cm. Once this charge reaches the anode, it is read out by an array of 120 charge tiles (10 $\times$ 10\,cm) consisting of perpendicular x and y strips with a pitch of 6\,mm~\cite{Li2019}. Using the fast timing information provided by the scintillation light, along with the location of hits on the charge tile, three-dimensional reconstruction of the incident particle interactions is possible. 

While the supernova neutrino detection trigger configuration is still under development, this study uses a conservative scintillation light threshold based on the number of collected OPs in the xenon volume. The collected number of OPs are a function of the number of produced OPs and the light collection efficiency (see Ref.~\cite{Adhikari2022} for details). Our chosen threshold corresponds to $\sim$100\% efficiency for 500\,keV depositions. See Sec.~\ref{Sec:lightReconstruction} for more details on the example reconstruction algorithm used in this analysis and this study's trigger threshold. The large number of expected electron-antineutrino interactions in the active water shield could also be used to trigger the TPC data acquisition to search for lower-energy depositions in the xenon~\cite{AlKharusi2021}.

Steady-state backgrounds are not a major concern for supernova detection in nEXO, owing to the large energy of \nueXeCC{} events, low muon flux, and short duration of the CCSN burst. Potential sources of background are pileup of lower-energy events, and muons that do not trigger the veto system. For this study, we assume these backgrounds are negligible. The contribution from them and their expected contribution over the $\sim$10-sec CCSN burst could be studied with steady-state data collected during nEXO's planned 10-yr exposure. 

%%%%%%%%%%%%%%%%%%%%%%%%%%%%%%%%%%%%%%%%%%%%%%%%%%%%%%%%%%%%%%%%%%%%%%%%%%%%%%%%%%%%%%%%%%%%%
\section{Modeling xenon charged-current interactions with MARLEY}\label{Sect:MARLEY}
As the first step toward simulating electron-neutrino CC interactions with xenon nuclei in nEXO, the particles produced from these interactions are modeled using the MARLEY event generator~\cite{MARLEY}. These reactions as described by the following equation,
\begin{equation}
 \nu_e + {}^{134,136}\mbox{Xe}\rightarrow e^{-} + {}^{134,136}\mbox{Cs}^*
 \label{Eq:xeCC}
\end{equation}
where the asterisk indicates that the resulting cesium nucleus is typically expected to be generated in an excited state, which can subsequently deexcite through the emission of $\gamma$ rays, neutrons, and other particles. MARLEY models CC interactions in the allowed approximation (four-momentum transfer $q \rightarrow 0$, Fermi motion neglected); details on the underlying physics models used by MARLEY can be found in Ref.~\cite{Gardiner2021}.

While originally designed for CC interactions on argon~\cite{MARLEYArgon}, MARLEY has been adapted for a variety of different nuclear targets~\cite{An2023_1,An2023_2}. The required inputs for MARLEY are the Gamow-Teller (GT) and Fermi (F) strength distributions. The former can be measured in charge-exchange reactions, such as $(p,n)$ and $({}^3\mbox{He},t)$, or can be calculated with a variety of theoretical approaches. 

MARLEY uses the supplied matrix element and an incident neutrino flux to calculate the cross section as well as the particles produced by CC interactions along with their energy and momenta.

\subsection{Gamow-Teller and Fermi strengths}
For the simulations within this paper, experimentally measured GT distributions are obtained for the ${}^{136}$Xe$\rightarrow^{136}$Cs transition from Ref.~\cite{Frekers2018} up to 4.5\,MeV, supplemented with a theoretical calculation of the discrete GT strength at higher energies from Ref.~\cite{Moreno2006}. For ${}^{134}$Xe, no experimental measurements of the GT strength distribution exist, so these are derived entirely from the theoretical calculations in Ref.~\cite{Moreno2006}. There are other theoretical calculations of the GT strength for the ${}^{136}$Xe$\rightarrow{}^{136}$Cs transition (such as Ref.~\cite{Terasaki2023}), but studying the impact of different nuclear physics models is beyond the scope of this work. These uncertainties have been shown to affect the reliability of reconstructing incident supernova neutrino flux parameters for argon-based detectors in Ref.~\cite{Abud2023}.

The experimentally measured GT matrix elements are multiplied by a factor of $g_A^2=1.26^2$ to form the weak matrix elements, adopting the MARLEY input format, with the value of $g_A$ determined from the normalization assumed in Ref.~\cite{Taddeucci1987}. While recent \textit{ab initio} calculations of matrix elements are able to reproduce $\beta$-decay rates~\cite{Gysbers2019}, calculations with approximate nuclear models, such as those used within our study, tend to overpredict the GT strength~\cite{Towner1987}. We multiply these theoretical matrix elements by a quenching factor corresponding to $g_{A,\mbox{eff}}^2=(0.7)^2$. The quenched value of $g_A$ for these calculated matrix elements is similar to what is used in Ref.~\cite{Moreno2006}, and was chosen to compare MARLEY's predictions to the calculations of Ref.~\cite{Pirinen2019}. Additionally, this value is in agreement with studies of $g_A$ quenching for targets in a similar mass range~\cite{Pirinen2015}. 

The Fermi strengths assume a value of $B(F)=g_V^2(N-Z)$ where $N$ is the number of neutrons in the target nucleus, $Z$ is the number of protons, and $g_V=1$ is the vector coupling constant. This Fermi strength is assumed to be located entirely at the isobaric analog state (IAS) of the product nucleus, which is typical for most nuclei with $N\gg{}Z$~\cite{Frekers2018}. For ${}^{136}$Xe this IAS has been measured to occur at an excitation energy of 13.386\,MeV~\cite{Frekers2018} relative to the ${}^{136}$Cs ground state. For ${}^{134}$Xe, the formalism in Ref.~\cite{Frekers2018}, Eq.~\ref{Eq:IAS}, is used to predict the location of this state, $E_X^{IAS}$, which is estimated to be accurate to within a few hundred keV~\cite{Frekers2018},
%IAS equation
\begin{equation}
	\begin{aligned}
		&E_X^{IAS} = \Delta E_C + M(A,Z) - M(A,Z+1) + M(H) - M(n) \\
		&\Delta E_c = 1.44\left(Z+\frac{1}{2}\right)A^{-1/3} - 1.13 \mbox{ (MeV).}
	\end{aligned}
	\label{Eq:IAS}
\end{equation}
Here $M(Z,A)$ and $M(Z+1,A)$ refer to the atomic masses of the initial and final nucleus, $M(n)$ is the mass of the neutron, and $M(H)$ is the mass of a hydrogen atom. The calculated location of the isobaric analog state in ${}^{134}$Cs is 12.189\,MeV, using the mass evaluation from Refs.~\cite{Huang2021,Wang2021}. As a check of the validity of this equation in this mass range, we calculate the value of the IAS of ${}^{136}$Xe in ${}^{136}$Cs via Eq.~\ref{Eq:IAS} and obtain a value of 13.258\,MeV, in good agreement with the measured value. The measured location is used in the MARLEY input file.

For this study, we focus on the isotopes of primary interest to the nEXO detector. This work could be extended to cover other stable naturally occurring isotopes of xenon if experimental or theoretical Gamow-Teller strength predictions were readily available, particularly for stable non even-even isotopes.

\subsection{Deexcitation data}
MARLEY utilizes deexcitation data from \texttt{TALYS}\,1.6~\cite{talys} to predict the observed discrete $\gamma$'s emitted from a \nueXeCC{} interaction. While data exist for ${}^{134}$Cs excited states, until recently there were little experimental data on the excited states of ${}^{136}$Cs. Data from Ref.~\cite{Rebeiro2023,rebeiro2018} are added to the deexcitation data MARLEY uses; data from Ref.~\cite{Frekers2018} are used for higher-energy discrete states. We follow the approach from \texttt{TALYS} for spin-parity assignment when those quantities are unknown for a specific excited state~\cite{talysmanual}. To assign spin, a histogram is generated from the spins of lower-energy states and compared to a Wigner distribution, and spin is assigned based on the bin that most underestimates the Wigner distribution, using a spin cutoff parameter described by Eq. 237 within Ref.~\cite{talysmanual}. To assign parity, the distribution of parities of states below the current level is generated, and parity is assigned to balance that distribution. The MARLEY approach of using a standard Lorentzian to model deexcitations is followed to generate predictions for branching ratios~\cite{Gardiner2021}. Improved measurements of the spin-parity of these excited states and their branching ratios would lead to more accurate predictions of experimental signatures in the future, although this is expected to have a small impact on the predicted energy distribution from \nueXeCC{} interactions in nEXO.

\subsection{Other neutrino interactions in nEXO}
While the focus of this paper is on \nueXeCC{}, supernova neutrinos can also interact with xenon in nEXO through several other channels. We discuss CE$\nu$NS, inelastic NC neutrino-nucleus scattering, neutrino-electron elastic scattering  (\nuES{}), and electron-antineutrino charged-current scattering ($\bar{\nu}_e$CC-Xe) as sources of signal and background below.

The large CE$\nu$NS cross section, proportional to the number of neutrons in the target nucleus squared, combined with sensitivity to all flavors of supernova neutrino, results in a substantial number of expected CE$\nu$NS interactions in nEXO. However, the only signature of a CE$\nu$NS interaction is a low-energy (keV-scale) quenched nuclear recoil, which by itself would be insufficient to generate a trigger in nEXO. It has been suggested that CE$\nu$NS may be observable through an increase in single photoelectrons (the ``CE$\nu$NS glow''~\cite{Scholberg2019,Major2021}), but this has not yet been studied for nEXO. 

Inelastic NC neutrino-nucleus scattering can produce MeV-scale $\gamma$'s in nEXO and is sensitive to all flavors of supernova neutrinos. However, this process is predicted to have a smaller cross section than that of \nueXeCC{}~\cite{Pirinen2018}. Modeling these types of interactions in nEXO is beyond the scope of this paper, and could be a potential future area of study.

\nuES{}, described by Eq.~\ref{Eq:veES}, 
\begin{equation}
 \nu + e^- \rightarrow \nu^{\prime} + e^{-\prime},
 \label{Eq:veES}
\end{equation}
can produce MeV-scale signals, but is expected to have a smaller cross section than that of \nueXeCC{}. The total cross section for this process depends on the flavor of the interacting neutrino, with the largest corresponding to $\nu_e$s. We have included \veES{} elastic scattering in our simulations, but have not included contributions from $\bar{\nu}_e$s and $\nu_x$. The \nuES{} cross section for $\bar{\nu}_e$ is smaller by a factor of  $\sim$2, and for $\nu_x$ it is smaller by a factor of $\sim$4.

Electron antineutrinos can also interact through charged-current scattering on xenon, although the cross section for this interaction is expected to be approximately 2 orders of magnitude lower than that of electron neutrinos~\cite{Pirinen2019}, as a result of Pauli blocking~\cite{Chauhan2017}.

Finally, as discussed in Ref.~\cite{Haselschwardt2020}, it may be possible to tag \nueXeCC{} interactions through timing coincidences with short-lived metastable states in the resulting cesium nucleus, such as those recently identified for ${}^{136}$Cs~\cite{Rebeiro2023,Haselschwardt2023}. This would enable separating  \nueXeCC{} events from other types of neutrino interactions. This is not incorporated into our simulation, and may be difficult at supernova neutrino energies owing to the small metastable deexcitation energy relative to the prompt energy deposition.

\subsection{Comparison of inclusive cross section predictions}\label{Sect:comparison}
\begin{figure}[ht]
	\centering
	\includegraphics[width=\linewidth]{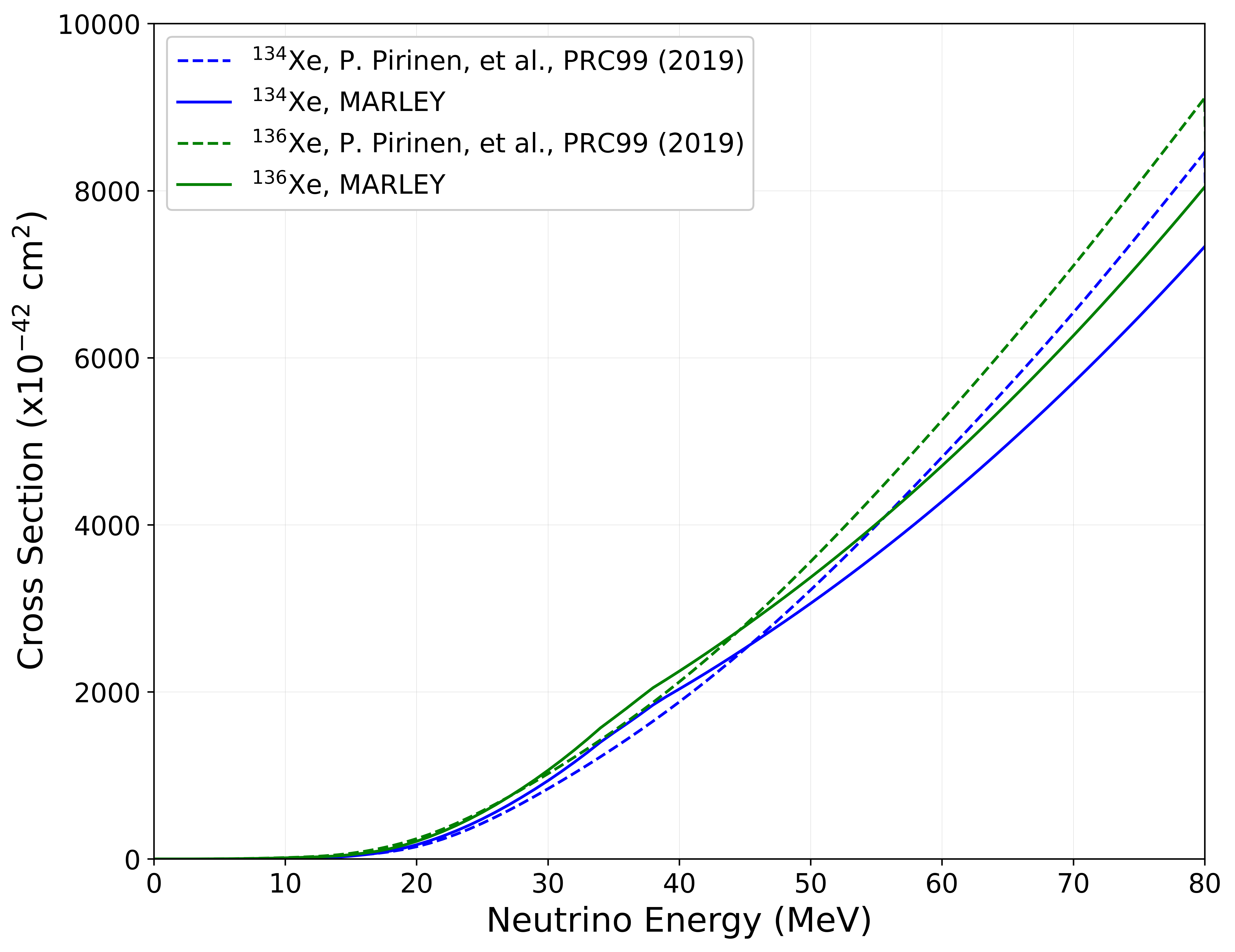}
	\caption{Comparison of MARLEY's prediction for the inclusive electron-neutrino CC cross section (using GT strengths from Refs.~\cite{Frekers2018,Moreno2006}) to the calculations in Ref.~\cite{Pirinen2019}. Both the matrix elements used with MARLEY and those from Ref.~\cite{Pirinen2019} employ a quasiparticle random-phase approximation approach and have the same quenching applied for comparison, corresponding to $g_A=0.7$.}
	\label{Fig:xsections_pirinen}
\end{figure}

As a test of MARLEY's predictions with the supplied nuclear data, cross sections for $\nu_e$CC interactions with ${}^{134}$Xe and ${}^{136}$Xe from MARLEY are compared to the calculations in Ref.~\cite{Pirinen2019}, shown in Fig.~\ref{Fig:xsections_pirinen}. In that reference, calculations of CC cross sections for stable isotopes of xenon are provided, including contributions from forbidden transitions. As can be seen in the figure, MARLEY agrees fairly well with those predictions at lower incident neutrino energies, but its predictions are smaller at higher energies. This could be, in part, due to the omission of forbidden transitions in the approximation used by MARLEY. In Appendix~\ref{appendixC} we calculate the inclusive cross section for ${}^{132}$Xe as a comparison.

\begin{figure}[ht!]
	\centering
	\includegraphics[width=\linewidth]{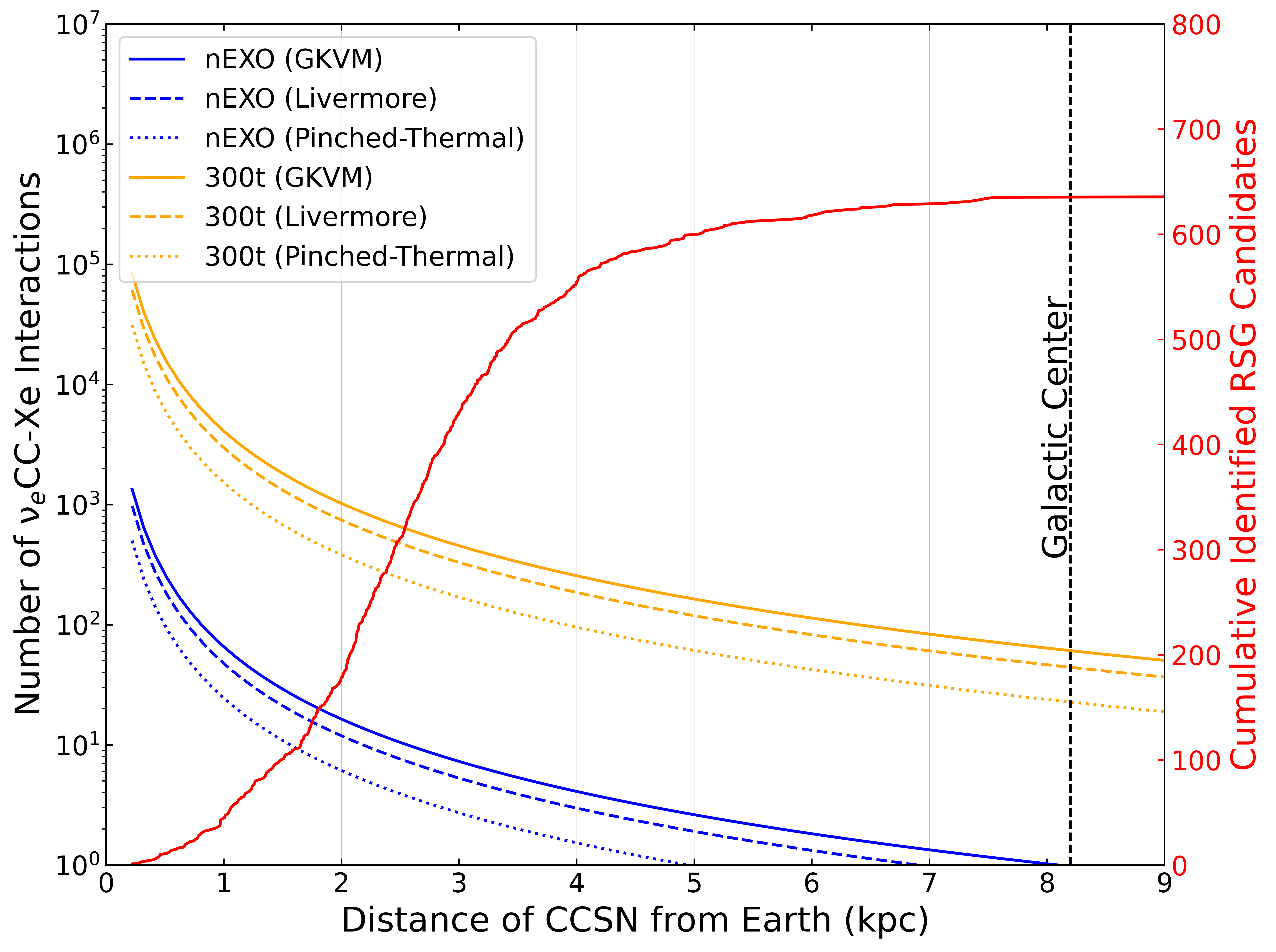}
	\caption{The number of supernova electron-neutrino interactions (\XeFCC{}, \XeSCC{} and \veES{}) occurring in nEXO and a 300-metric-ton detector of the same enrichment as a function of distance. No detector efficiencies are taken into account. The red line shows the cumulative number of candidate RSGs from Ref.~\cite{Healy2024}. nEXO's expected range extends out to approximately the Galactic Center (black dashed line), which covers the majority of the identified RSG candidates in that survey, although the expected number in our Galaxy is much larger~\cite{Gehrz1989}. The supernova flux models correspond to those from Refs.~\cite{gkvm,livermore}, and a pinched-thermal spectrum as described in Sec.~\ref{Sec:pinchedThermal}.}
	\label{Fig:interactionRate}
\end{figure}

The \nueXeCC{} flux-averaged cross sections are computed using three commonly used neutrino flux models: the Gava-Kneller-Volpe-McLaughlin (GKVM) model~\cite{gkvm} and the Livermore model~\cite{livermore}, both from Ref.~\cite{snowglobes}, and a ``pinched-thermal" spectrum with parameters $(\alpha,\varepsilon,\langle E_{\nu}\rangle)$ of (2.5, 5$\times 10^{52}$\,erg, 9.5\,MeV), from Ref.~\cite{Abi2021}. The flux-averaged cross sections from MARLEY agree well with the predictions from Ref.~\cite{Pirinen2019}, as can be observed in Table~\ref{Tab:1}. While flux-averaged cross sections vary significantly with different neutrino flux models, the normalization for the electron-neutrino flux and average electron-neutrino energy differ as well, so the expected interaction rates with different models are closer together (see Fig.~\ref{Fig:interactionRate}).

\begin{table}[ht!]
\begin{center}
\begin{tabular}{cccc}
\hline\hline
\multicolumn{4}{c}{${}^{136}$Xe Cross Section ($\times 10^{-40}\mbox{cm}^2$)}\\
\hline
&GKVM~\cite{gkvm} & Livermore~\cite{livermore} & Pinched thermal \\
\hline
Ref.~\cite{Pirinen2019} & 3.15 & 0.89 & 0.43 \\
MARLEY & 3.10 & 0.83 & 0.38 \\
\hline
\multicolumn{4}{c}{${}^{134}$Xe Cross Section ($\times 10^{-40}\mbox{cm}^2$)}\\
\hline
Ref.~\cite{Pirinen2019} & 2.49 & 0.63 & 0.28 \\
MARLEY & 2.68 & 0.67 & 0.28 \\
\hline
\multicolumn{4}{c}{CCSN Neutrino Model Comparison}\\
\hline
$\nu_e$ $(\times 10^{57})$&1.16 & 3.06 & 3.29\\
$\langle E_{\nu_e} \rangle $ (MeV) & 16.5 & 11.3 & 9.5\\
\hline\hline
\end{tabular}
\caption{Comparison of the predictions for inclusive cross sections from MARLEY, using the matrix elements from Refs.~\cite{Frekers2018,Moreno2006}, to those from Ref.~\cite{Pirinen2019}. For the calculations from Ref.~\cite{Pirinen2019}, a spline interpolation is used to evaluate the cross section within the specified flux model from the interaction threshold up to 80\,MeV, whereas all MARLEY cross sections are evaluated up to 100\,MeV. The pinched-thermal cross section corresponds to $\alpha=2.5$, $\varepsilon=5 \times 10^{52}$\,erg, and $\langle E_{\nu} \rangle = 9.5$\,MeV, from Ref.~\cite{Abi2021}, as described in Sec.~\ref{Sec:pinchedThermal}. Also shown is the integrated number of $\nu_e$s produced over the 10-sec burst and average $\nu_e$ energy in the various models.}
\label{Tab:1}
\end{center}
\end{table}

Using the inclusive cross section predictions from MARLEY and an assumed xenon mass of 4,811\,kg (from Ref.~\cite{Adhikari2022}) enriched to 90\% in ${}^{136}$Xe, the predicted number of electron-neutrino interactions within nEXO is plotted as a function of distance in Fig.~\ref{Fig:interactionRate}, including contributions from both \nueXeCC{} and \veES{}. Also shown is the cumulative number of candidate red supergiants (RSGs), progenitors of supernovae, within our Galaxy as a function of distance from Earth, from Ref.~\cite{Healy2024}. As noted in Ref.~\cite{Healy2024}, the identified set of candidate RSGs is not complete, with an estimated $\sim$5,000 RSGs predicted to exist within our Galaxy~\cite{Gehrz1989}. nEXO's sensitivity extends roughly to the Galactic Center of the Milky Way. We calculate there will nominally be one or more electron-neutrino interactions in nEXO for CCSNe out to $\sim$5-8\,kpc, depending on the supernova neutrino flux model and parameters of the CCSN. Within that distance of Earth there are several hundred candidate RSGs that may produce neutrino-xenon interactions in nEXO. A 300-metric-ton detector of similar enrichment would be able to observe events out to significantly larger distances, with $\sim$15-41 events predicted at a distance of 10\,kpc depending on which supernova electron-neutrino flux model is used.

\subsection{Exclusive cross section predictions}
When a neutrino interacts with a nucleus, various particles (predominantly neutrons and $\gamma$ rays) can be emitted as a result of the deexcitation of the product nucleus. MARLEY generates predictions not only for inclusive cross sections, but also for the various exclusive cross sections and particle distributions resulting from \nueXeCC{} interactions. Exclusive cross sections leading to bound cesium states (no nucleon emission), along with those resulting in neutron emission, are shown in Table~\ref{Tab:2}. The dominant exclusive deexcitation channel is expected to be neutron emission, although recent measurements of CC neutrino interactions on ${}^{127}$I and Pb with $\sim$30\,MeV neutrinos have suggested this channel may be suppressed for heavy nuclei~\cite{An2023_1, An2023_2}. 

\begin{table*}[ht!]
	\begin{center}
		\hspace*{-0.7cm}\begin{tabular}{cccc}
			\hline\hline
			\multicolumn{4}{c}{${}^{136}$Xe ($\times 10^{-40}\mbox{cm}^2$)}\\
			\hline
			Channel & GKVM~\cite{gkvm} & Livermore~\cite{livermore} & Pinched thermal\\
			\hline
			${}^{136}$Xe($\nu_e,e^-)$&3.10&0.83&0.38\\
			${}^{136}$Xe($\nu_e,e^-){}^{136}\mbox{Cs}_{\mbox{\tiny bound}}$&0.64&0.29&0.18\\
			${}^{136}$Xe($\nu_e,e^-+n){}^{135}\mbox{Cs}$&2.44&0.55&0.20\\
			${}^{136}$Xe($\nu_e,e^-+2n){}^{134}\mbox{Cs}$&0.02&0.00&0.00\\
			\hline
			
			\multicolumn{4}{c}{${}^{134}$Xe ($\times 10^{-40}\mbox{cm}^2$)}\\
			\hline
			Channel & GKVM~\cite{gkvm} & Livermore~\cite{livermore} & Pinched thermal\\
			\hline
			${}^{134}$Xe($\nu_e,e^-)$&2.68&0.67&0.28\\
			${}^{134}$Xe($\nu_e,e^-){}^{134}\mbox{Cs}_{\mbox{\tiny bound}}$&0.57&0.21&0.12\\
			${}^{134}$Xe($\nu_e,e^-+n){}^{133}\mbox{Cs}$&2.10&0.46&0.16\\
			${}^{134}$Xe($\nu_e,e^-+2n){}^{132}\mbox{Cs}$&0.01&0.01&0.00\\
			\hline
			
			\multicolumn{4}{c}{$\nu_e-e^-$ ($\times 10^{-40}\mbox{cm}^2$)}\\
			\hline
			Channel & GKVM~\cite{gkvm} & Livermore~\cite{livermore} & Pinched thermal\\
			\hline
			$\nu_e-e^-$&0.08&0.06&0.05\\
			\hline\hline
		\end{tabular}
		\caption{Calculated exclusive electron-neutrino cross sections on xenon from MARLEY, using the same assumptions as described in the caption of Table~\ref{Tab:1}. For neutrino-electron elastic scattering, this cross section is given per xenon atom.}
		\label{Tab:2}
	\end{center}
\end{table*}

\subsection{Predicted particles and visible energy}
\begin{figure}[ht]
	\centering
	\includegraphics[width=\linewidth]{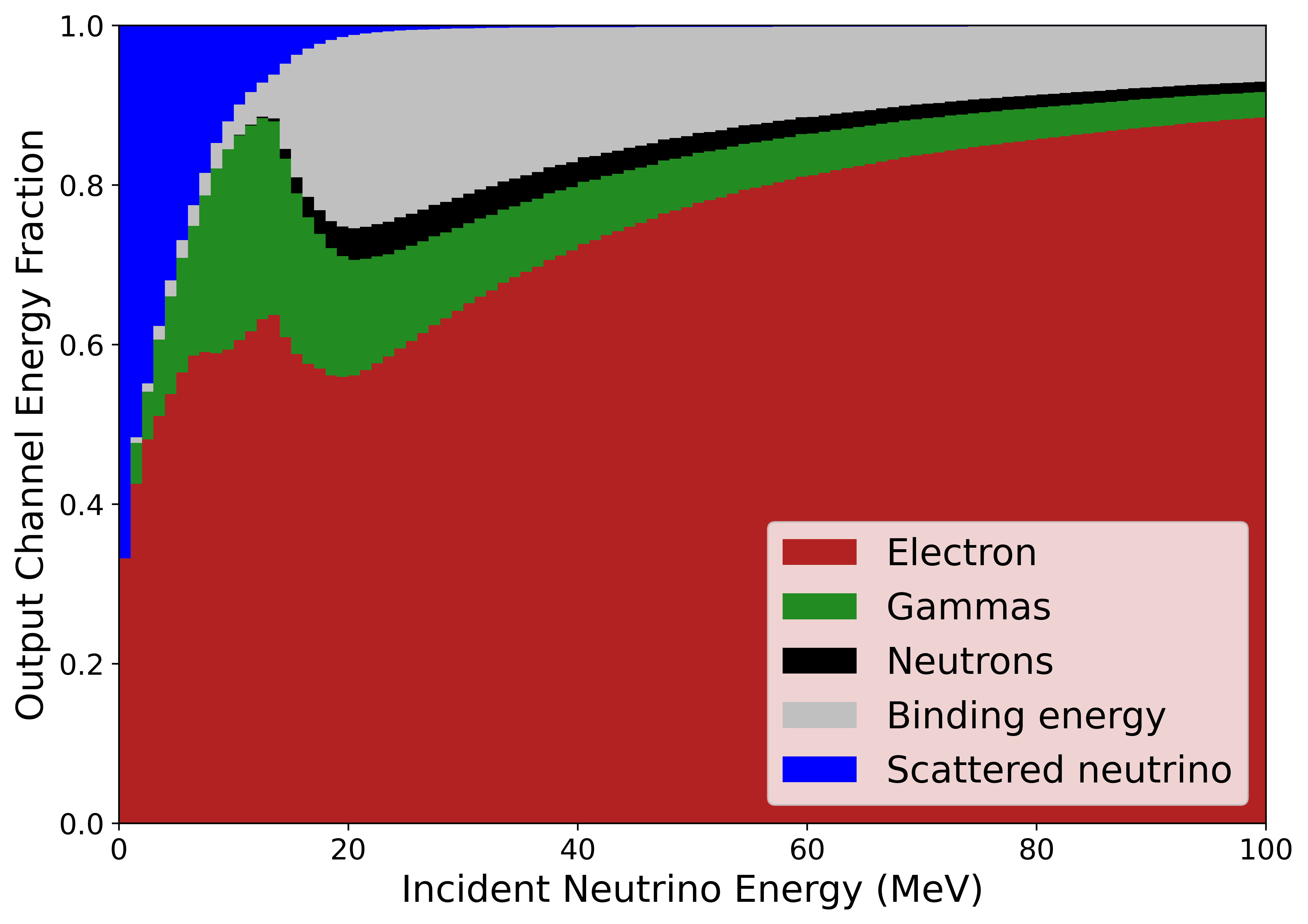}
	\caption{Predicted energy distribution of particles leaving an electron-neutrino interaction (\XeFCC{}, \XeSCC{} and \veES{}).}
	\label{Fig:distribution}
\end{figure}

Using MARLEY, the fraction of incident neutrino energy transferred to various output channels is visualized in Fig.~\ref{Fig:distribution}. This plot incorporates \XeFCC{}, \XeSCC, and \veES{} interactions, with appropriate cross section weighting and isotopic abundances (90\% ${}^{136}$Xe, 10\% ${}^{134}$Xe). Individual distributions for these three processes can be found in Appendix~\ref{appendixA}.

\begin{figure}[ht]
	\centering
	\includegraphics[width=\linewidth]{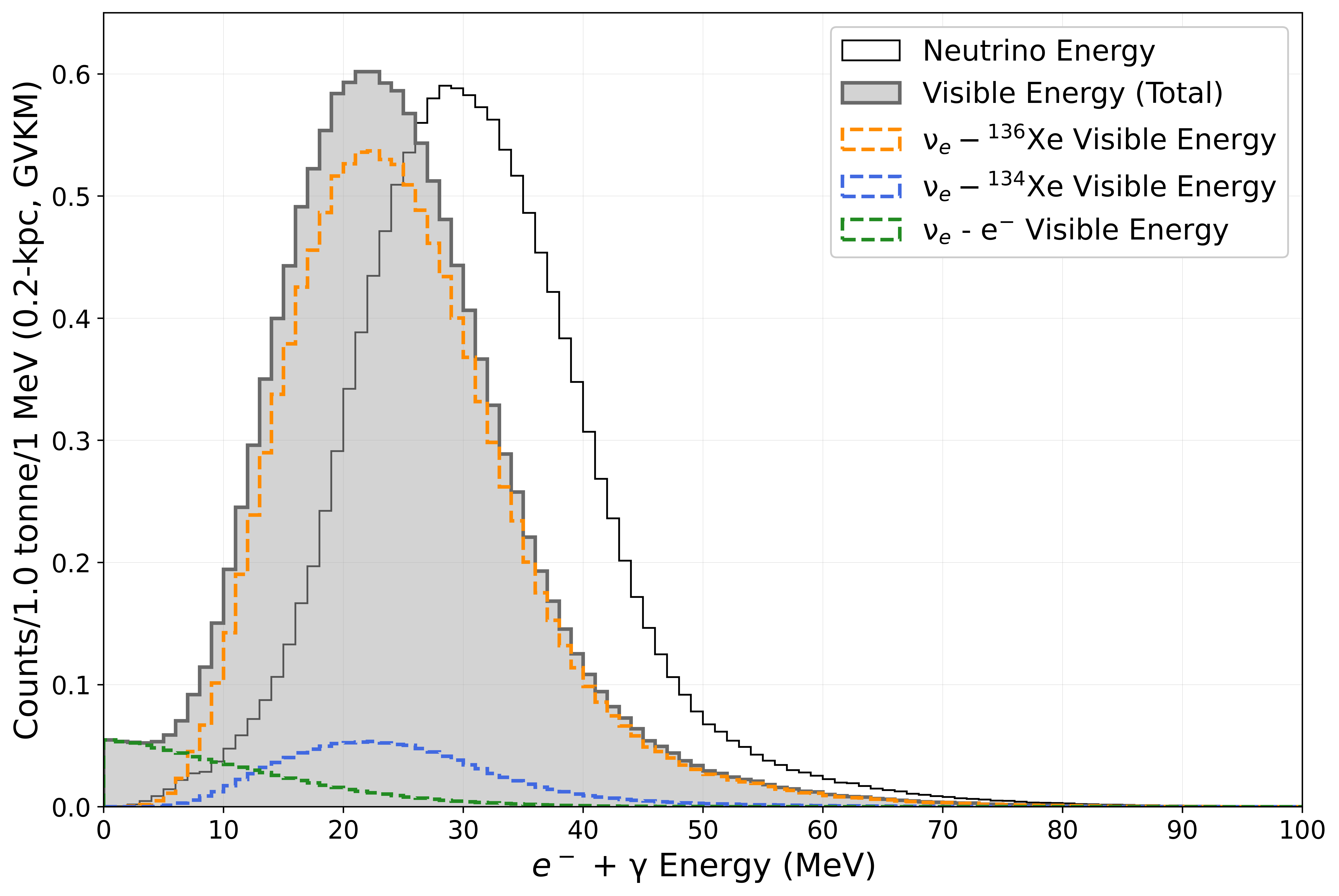}
	\caption{Comparison of the energy of interacting electron neutrinos and the corresponding visible energy spectrum using the GKVM flux model~\cite{gkvm}. Neutrino-electron elastic scattering from $\bar{\nu}_e$ and $\nu_x$ are not included in this plot.}
	\label{Fig:emEnergy}
\end{figure}

In the interactions modeled, some energy will be lost to nonvisible channels (thresholds for CC interactions, neutron-binding energy, and the scattered neutrino from elastic neutrino-electron scattering interactions). Additionally, emitted neutrons and the recoiling xenon nucleus also produce scintillation light and ionization charge, although these signals are quenched so their contribution is small~\cite{Lenardo2015}. To approximate the energy that would be visible in a detector from electron-neutrino interactions (neglecting detector thresholds, geometric size, and efficiencies), the distribution of visible (scintillation) energy is plotted along with the incident neutrino energy in Fig.~\ref{Fig:emEnergy}, using an incident electron-neutrino spectrum characterized by the GKVM supernova flux model~\cite{gkvm}. Here, visible energy ($E_{\mbox{vis}}$) is defined as the energy of resulting $\gamma$ rays and electrons,
\begin{equation}\label{Eq:Evis}
	E_{\mbox{vis}} = E_{e^-} + E_{\gamma}
\end{equation}
where the contribution from the recoiling nucleus as well as elastic and inelastic scattering of emitted neutrons is neglected, as a full simulation is required to accurately describe the energy deposited from these interactions (Sec.~\ref{Sect:reconstruction}).

%%%%%%%%%%%%%%%%%%%%%%%%%%%%%%%%%%%%%%%%%%%%%%%%%%%%%%%%%%%%%%%%%%%%%%%%%%%%%%%%%%%%%%%%%%%%%
\section{Simulation and event reconstruction}\label{Sect:reconstruction}
Electron-neutrino interaction events were simulated using Geant4 v10.7.2~\cite{geant_1,geant_2,geant_3} along with a modified version of the \texttt{Noble Element Simulation Technique}\,2.0.1~\cite{nest,nest_2}. For more details on the simulation framework, refer to Ref.~\cite{Adhikari2022}.

Compared to $0\nu\beta\beta$ and most backgrounds of primary interest for nEXO, the predicted signals from supernova electron-neutrino interactions are higher in energy, and can have delayed components due to neutron emission and subsequent capture. An example reconstruction algorithm was developed based of the reconstruction described in Ref.~\cite{Adhikari2022}, modified for this analysis. We use only scintillation light as our energy estimator in this analysis for two reasons. 

First, there is less ambiguity in distinguishing prompt electromagnetic depositions from delayed neutron capture events in the scintillation channel based on timing, whereas this can be more difficult for the ionization channel due to the long drift times (potentially hundreds of microseconds) and the fact that neutrons can travel a significant distance prior to capturing. 

Second, the use of charge as an energy estimator would require a stricter fiducial-volume cut to select only charge events fully contained in the TPC drift region. As multiple MeV-scale $\gamma$'s can be emitted by a CC event as part of the cesium deexcitation process, CC events are spatially larger than the events typically of interest to nEXO. Using light as the energy estimator allows for the inclusion of events that extend into the xenon space outside the drift region (referred to as the xenon skin). 

While we use light as our energy estimator, we reconstruct the position and magnitude of charge depositions to use a simulated map of the geometric dependence of light collection efficiency across the detector (see Ref.~\cite{Adhikari2022}) to correct for the position dependence of light collection. The impact of both grouping a light signal with an incorrect ionization signal or using partially contained charge depositions to apply this correction is a small increase in the energy resolution of our reconstructed light signals using this algorithm. In the future, an improved charge timing and energy reconstruction can be developed for nEXO supernova events, though for this study we choose to start with the above simplifications.

\subsection{Photon reconstruction}\label{Sec:lightReconstruction}
For scintillation light reconstruction, the number of collected OPs was calculated based on the true interaction position, the number of produced photons, a map of the simulated geometric dependence of the photon collection efficiency, the photon detection efficiency of the SiPMs, and the correlated avalanche probability (based on values measured for prototype nEXO SiPMs~\cite{Adhikari2022,Gallina2022}). As in Ref.~\cite{Adhikari2022}, dark counts and electronic noise were neglected due to their small expected contribution~\cite{Jamil2018,Gallina2019}. The number of OPs was calculated in a moving 300-ns window (approximately 10 times the mean decay time of the long component of xenon scintillation~\cite{Abe2018}), and if the sum in that window exceeded a threshold (in terms of collected OPs, corresponding to $\sim$100\% efficiency for 500 keV events within the drift region), the integrated number of OPs within that window was recorded. 

SiPM saturation was not included in the reconstruction algorithm, given the large spatial distribution of the energy depositions in CC events and isotropy of scintillation light production. The nEXO photon readout electronics are set to clip SiPM channels with more than 100 collected photoelectrons (PEs). A subset of \nueXeCC{} events were simulated using \texttt{CHROMA}~\cite{chroma1,chroma2} to study collected photon distributions in SiPM channels. While 55.6\% of reconstructed supernova neutrino events had at least one SiPM channel collecting more than 100\,PEs, for events where clipping was present the mean fraction of clipped SiPM channels was 1.67\%. For these events, nonclipped quantities, such as the tail integral~\cite{Pershing2022} or nonsaturated SiPM channels, can be used to estimate event energy at the cost of an increase in energy resolution. Additional work is needed to quantify the impact of these mitigating approaches on energy resolution, and we have not included this effect in our study. 

\subsection{Charge reconstruction}
The number of simulated drifted ionization electrons collected by charge tiles is calculated, including effects of electron lifetime attenuation as well as longitudinal and transverse dispersion. Similar to the light reconstruction algorithm, a rolling window of 120\,$\mu$s is used to sum the number of detected electrons in each individual charge channel; if these exceed the threshold of 500 electrons, approximately 2.5 times the rms channel noise, the integrated charge in that channel is recorded. 

Noise is added to the integrated charge on each channel based on a distribution derived using an optimum filter~\cite{Smith1997,Gowala2000} for reconstruction based on an internal nEXO collaboration study. This was found to be in good agreement with the data, and produces an energy resolution near the $0\nu\beta\beta$ $Q$ value consistent with that of Ref.~\cite{Adhikari2022}. A small number of events produce charge on channels that exceed the 120,000 electrons used in that study, so a $1/A$ dependency was fit to the channel-integrated charge noise for lower-energy charge deposits and used to extrapolate noise for higher single-channel charge deposits, where $A$ is the integral of the pulse. 

The potential of having multiple clusters of charge deposition (e.g. prompt $\gamma$ ray and electron depositions plus delayed neutron captures) separated in both space and time drives the need for a clustering algorithm to group charge depositions correlated with a photon signal. The input to the clustering algorithm were charge-weighted points, $Q_{x_i,y_j}$, located at the intersections of x and y strips on each charge tile,
\begin{equation}
	\label{Eq:chargeDist}
	Q_{x_i,y_j} = Q_{x_i}\frac{Q_{y_j}}{Q_{y_{tile}}} + Q_{y_j}\frac{Q_{x_i}}{Q_{x_{tile}}} 
\end{equation}
where $Q_{x_i}$ is the charge on channel $x_i$, $Q_{y_j}$ is the charge on channel $y_j$, and $Q_{x_{tile}}$ and $Q_{y_{tile}}$ are the sum of the charge on the x and y channels of the tile respectively. Then, the \texttt{DBSCAN} algorithm~\cite{dbscan,dbscan2} implementation of \texttt{scikit-learn}~\cite{scikit} is used to group individual charge points to form charge clusters. The \texttt{DBSCAN} algorithm has two free parameters: \texttt{eps}, which corresponds to the maximum distance between two points in the cluster for them to be considered within the same neighborhood, and \texttt{min}\_\texttt{samples}, which corresponds to the minimum number of samples in a neighborhood to consider a point to be a core point. \texttt{min}\_\texttt{samples} was left at its default value of 5, and \texttt{eps} was set to a value of 1,000\,mm for this analysis. This is a conservative value chosen empirically by comparing the number of reconstructed charge clusters to the number of reconstructed light clusters. The purpose of the clustering algorithm is not to identify individual particles produced by electron-neutrino interactions, but instead to group interactions as part of either the prompt signal or the delayed neutron capture. The optimized large value likely originates from the broad spatial distribution of the emitted $\gamma$ rays and electron from these events, as well as scatters of the emitted neutron that can occur far from the electron-neutrino interaction location. 

Charge-channel saturation was not incorporated into the reconstruction algorithm. Assuming a 12-bit digitizer is used for the charge readout, a baseline located at one-third of the digitizer range, and a gain setting corresponding to $\sim$43\,$e$/analog-to-digital converter counts, it was determined that $\sim$0.04\% of events have at least one charge channel that would be saturated. The CC events simulated are high in energy, but typically have broad spatial distributions, owing to the production of a number of deexcitation $\gamma$'s, so the saturation of an individual charge channel occurs infrequently. Additionally, as with light signals, it may be possible to use unsaturated portions of charge signals when saturation is present. These assumptions are motivated by simulations of the optimal settings for the CRYO ASIC charge digitization chip planned for use in nEXO~\cite{PenaPerez2020,PenaPerez2022}. 

\subsection{Event formation}\label{Sec:LightCal}
After reconstructing light and charge, these events were correlated with each other. The approach taken was to form a list of all possible charge clusters that reconstruct to valid positions within the TPC when assuming correlation with the prompt light signal and to do the same for the delayed light signal (due to neutron capture) if present. A charge cluster that reconstructed a valid interaction position for only one of the light signals was assumed to be correctly associated and eliminated as a candidate charge event for the other light signal. If degeneracies were still present, the largest charge signal was assumed to be correlated with the largest light signal, which has an impact on the reconstructed energy resolution.

Using the location of the individual depositions within a charge cluster and a map of the geometric dependence of nEXO's photon collection efficiencies, a weighted-average light collection efficiency was determined for each event proportional to the charge at each x- and y-channel intersection point. This efficiency is used to estimate the true number of photons produced by the scintillation signal and thus the calibrated energy of the light signal.

This example algorithm was found to reconstruct electron-neutrino interactions generated with a pinched-thermal spectrum described above with 88.4\% efficiency. The remaining events were either missing a charge signal (no depositions within the drift region or above the charge-channel threshold) or a light signal (no depositions above scintillation light threshold). While our reconstruction algorithm is not able to obtain an energy estimate for this $\sim$11.6\% of events, it may still be possible to use them to further understand supernova electron-neutrino neutrino interactions.

Using the reconstructed events, we can tag neutron captures by looking for the presence of multiple light signals separated in time by at least 300\,ns. Neutron captures were observed for $\sim$33.6\% of the true neutron-emitting events. The identification of neutron-emitting events could potentially be improved by developing an algorithm based on event topology or using machine learning.

%%%%%%%%%%%%%%%%%%%%%%%%%%%%%%%%%%%%%%%%%%%%%%%%%%%%%%%%%%%%%%%%%%%%%%%%%%%%%%%%%%%%%%%%%%%%%
\section{Reconstructing the supernova neutrino spectrum}\label{Sect:parameterization}
Neutrinos of different energies and flavors decouple from thermal equilibrium at different depths in a CCSN. Thus, reconstructing the neutrino energy spectrum of different flavors allow for imaging of the interior dynamics of the collapse. Many factors can impact a detector's ability to reconstruct the incident neutrino spectrum, such as statistics, energy resolution, and detection threshold. There are a number of existing studies~\cite{Abud2023,Abi2021,Nikrant2018,Rosso2018,Minakata2008,Botella2004} evaluating the ability of different detectors to reconstruct an incident neutrino spectrum given an observed detector energy spectrum, but none of these focus on xenon-based detectors using the charged-current channel.

\subsection{Pinched-thermal spectrum}\label{Sec:pinchedThermal}
%Oscillations, neutrino mixing, MSW
A complete description of the expected supernova neutrino signal is complex, as it is affected by many factors, such as the neutrino mass splitting, self-induced flavor conversion, progenitor mass, and the Mikheyev–Smirnov–Wolfenstein effect~\cite{MSW}. While a complete description of the expected neutrino spectrum from a CCSN is beyond the scope of this study, more detail on the impact of these factors on the neutrino spectrum can be found in Refs.~\cite{Mirizzi2016,Saez2024}. Here we adopt a simplified pinched-thermal neutrino spectrum~\cite{Tamborra2012,Minakata2008}, 
\begin{align}
	&\phi(E_{\nu},\varepsilon,\langle E_{\nu} \rangle, \alpha, d) = \nonumber\\
	&\quad \frac{\varepsilon}{4\pi d^2} \frac{(\alpha+1)^{\alpha + 1}}{\langle E_{\nu}\rangle^2\Gamma(\alpha+1)} \left(\frac{E_{\nu}}{\langle E_{\nu} \rangle}\right)^{\alpha}e^{-(\alpha+1)E_{\nu}/\langle E_{\nu}\rangle}
 \label{Eq:pinchedThermal}
\end{align}
where $d$ is the distance to the collapsing star, $\varepsilon$ is the total binding energy released in the form of neutrinos by the CCSN, $\alpha$ is a positive-definite pinching parameter representing the deviation from a Fermi-Dirac spectrum, and $\langle E_{\nu} \rangle$ is the mean energy of the neutrinos. While there are many sets of parameters that may describe the spectrum of electron neutrinos incident on an Earth-based detector integrated over the $\sim$10-sec CCSN burst, we use the same example parameters as in Ref.~\cite{Abi2021} to allow for a comparison of the results within that reference. The example CCSN parameters chosen in that reference are $(\alpha, \varepsilon, \langle E_{\nu} \rangle) = (2.5, 5\times 10^{52}\mbox{ erg}, 9.5\mbox{ MeV})$. 

\begin{figure}[ht]
	\centering
	\includegraphics[width=\linewidth]{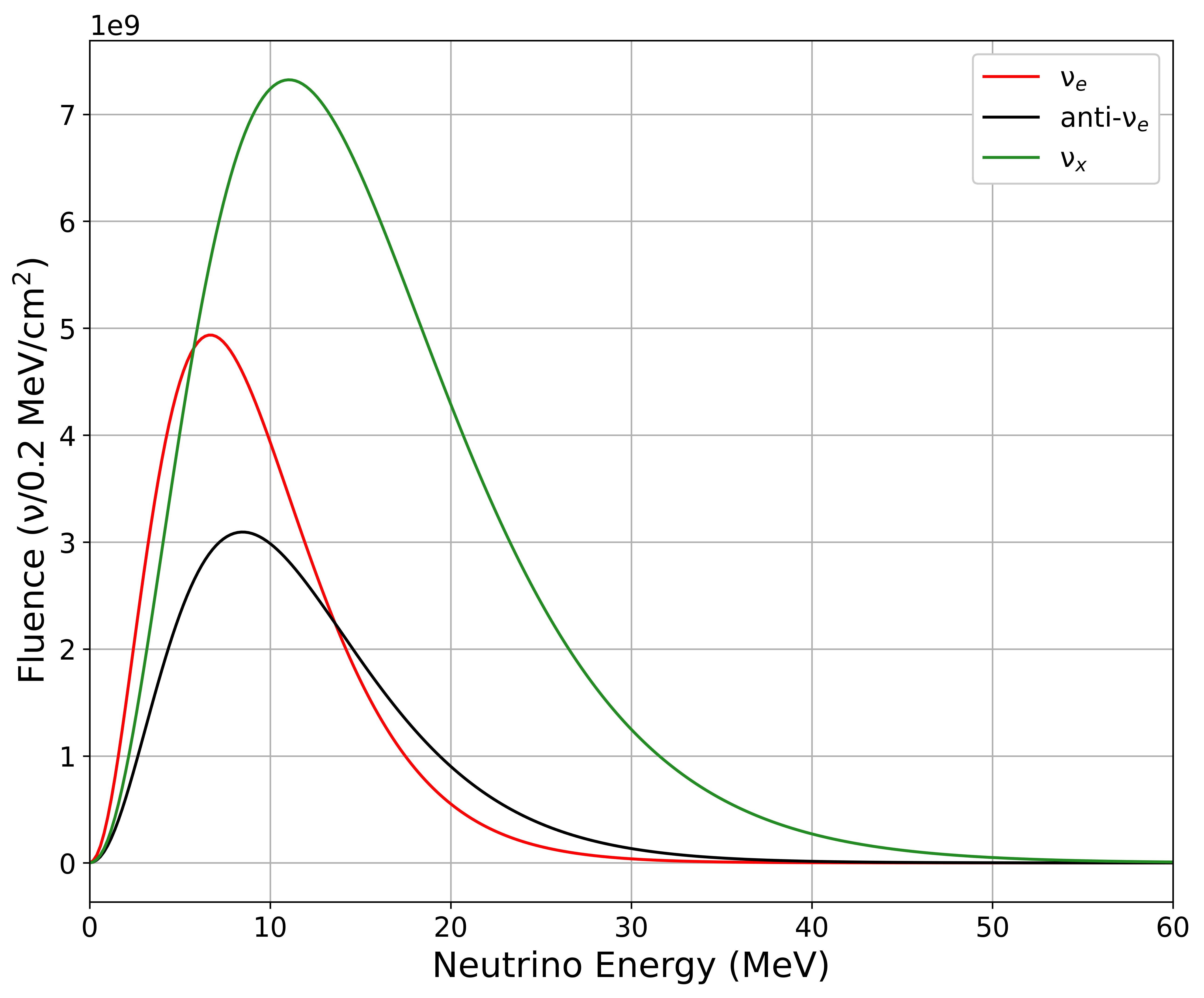}
	\caption{Pinched-thermal spectrum with $(\alpha, \varepsilon, \langle E_{\nu} \rangle) = (2.5, 5 \times 10^{52}\mbox{ erg}, 9.5\mbox{ MeV})$ for $\nu_e$, $(2.5, 5 \times 10^{52}\mbox{ erg}, 12\mbox{ MeV})$ for $\bar{\nu_e}$, and $(2.5, 5 \times 10^{52}\mbox{ erg}, 15.6\mbox{ MeV})$ for each flavor in the $\nu_x$ component.}
		\label{Fig:spectrum}
\end{figure}

There are a variety of other models formulated to describe the expected neutrino spectrum (such as Refs.~\cite{gkvm,livermore,garching}), but here we focus on only the pinched-thermal form as it is easy to parameterize. In principle, this work could be extended to other models using the transfer matrix we develop (Sec.~\ref{Sec:transferMatrix}). We also assume the distance to the collapsing star is well known, and fix this distance parameter in describing the spectrum. While not of relevance for this study, the electron-antineutrino component and NC component ($\nu_{\mu} + \nu_{\tau} + \bar{\nu_{\mu}} + \bar{\nu_{\tau}}$) can also be described by a pinched-thermal form, with a different set of parameters. Figure~\ref{Fig:spectrum} depicts these three components with parameter assumptions from Ref.~\cite{Abi2021}.

\subsection{Transfer matrix}\label{Sec:transferMatrix}
Using the methods described in the above text, $1 \times 10^6$ \XeFCC{}, \XeSCC{}, and \veES{} were each generated with MARLEY spanning a uniform energy distribution from 0 to 100\,MeV, simulated using the nEXO simulation framework, and analyzed with the previously discussed reconstruction algorithm. A transfer matrix was formed for each interaction type mapping incident neutrino energy into reconstructed detector energy. In the analysis, the individual transfer matrices from these three components were used, but to illustrate the total detector response these are combined with cross section and isotopic abundance weighting in Fig.~\ref{Fig:combined_transferMatrix}. The individual transfer matrices for each component can be found in Appendix~\ref{appendixB}. 

The total detector response is dominated by \XeSCC{} interactions, as CC cross sections are predicted to be larger than the elastic neutrino-electron scattering cross sections, and ${}^{136}$Xe will make up 90\% of the enriched xenon target. The source of the two main diagonal bands in Fig.~\ref{Fig:combined_transferMatrix} is the difference of binding energy for \XeSCC{} interactions with and without neutron emission, which are separated by 6.828\,MeV, as determined by the atomic mass evaluation of Refs.~\cite{Huang2021,Wang2021}.
 
While cuts on the data can improve the precision with which nEXO reconstructs the true energy deposited by a particle, leading to improved energy resolution, a minimal set of cuts were employed here to maximize statistics. A cut was developed to remove events with a reconstructed charge deposition center within 5\,cm of the nEXO cathode. This is greater than the standoff cut employed in the $0\nu\beta\beta$ analysis in part due to the large spatial size of these events leading to more energy deposited below the cathode where light collection efficiency is worse. Adopting this cut reduces the overall reconstruction efficiency from 88.4\% to 86.0\%. The transfer matrix utilized preserves the energy dependence and overall normalization of the reconstruction efficiency. 

\begin{figure}[ht!]
	\centering
	\includegraphics[width=\linewidth]{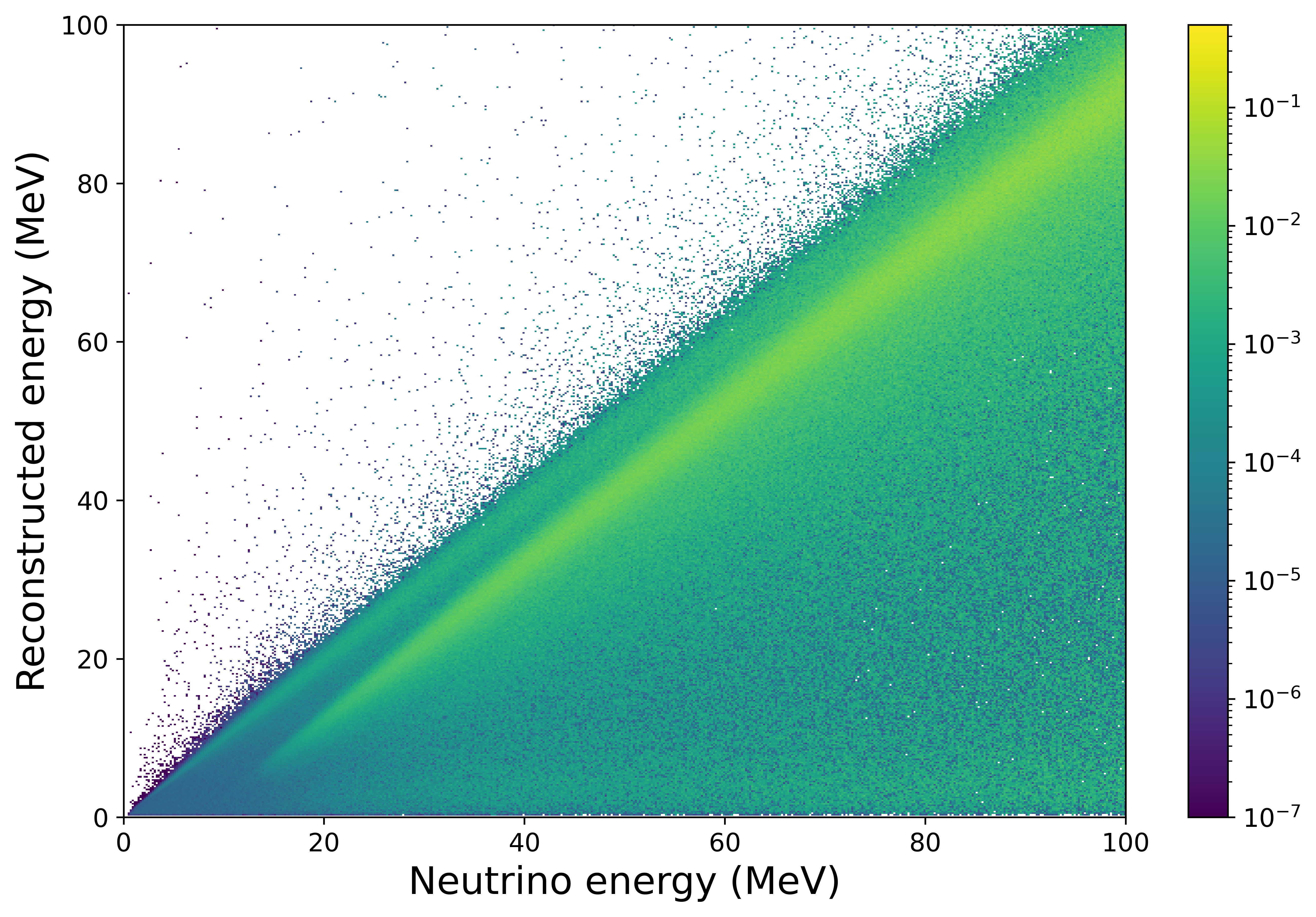}
	\caption{Transfer matrix for Xe interactions in nEXO, including \XeFCC{}, \XeSCC{}, and \veES{} interactions, mapping neutrino energy into reconstructed energy.}
	\label{Fig:combined_transferMatrix}
\end{figure}

\subsection{Fitting procedure}
The neutrino interaction events generated by MARLEY were run through the nEXO simulation framework and the described reconstruction algorithm. Using the reconstructed spectrum from the nominal set of pinched-thermal spectrum parameters, $1.5 \times 10^6$ datasets (referred to as test datasets) were generated with Poisson statistics for supernovae located at distances of 0.2 and 0.5\,kpc from Earth.

Using \texttt{RooFit}~\cite{roofit}, a likelihood function was developed to compare the simulated energy spectrum in the detector using a trial set of pinched-thermal parameters to the simulated energy spectrum of the test dataset. First, a trial pinched-thermal neutrino spectrum was generated and multiplied by the predicted cross sections from MARLEY for \XeFCC{}, \XeSCC{}, and \veES{} interactions. Next, the xenon mass and isotopic abundance of nEXO were used to generate the expected number of neutrino interactions of each type. Then, the transfer matrices
were used to incorporate detector response and generate extended probability distribution functions (PDFs) for each type of interaction, which were combined into a single extended PDF. The combined PDF had a bin size of 0.2\,MeV, and the test datasets fit to were unbinned. The likelihood function was then minimized using the \texttt{iminuit} python package~\cite{iminuit}. 

\subsection{Fit results}
\begin{figure}[ht!]
	\centering
	\includegraphics[width=\linewidth]{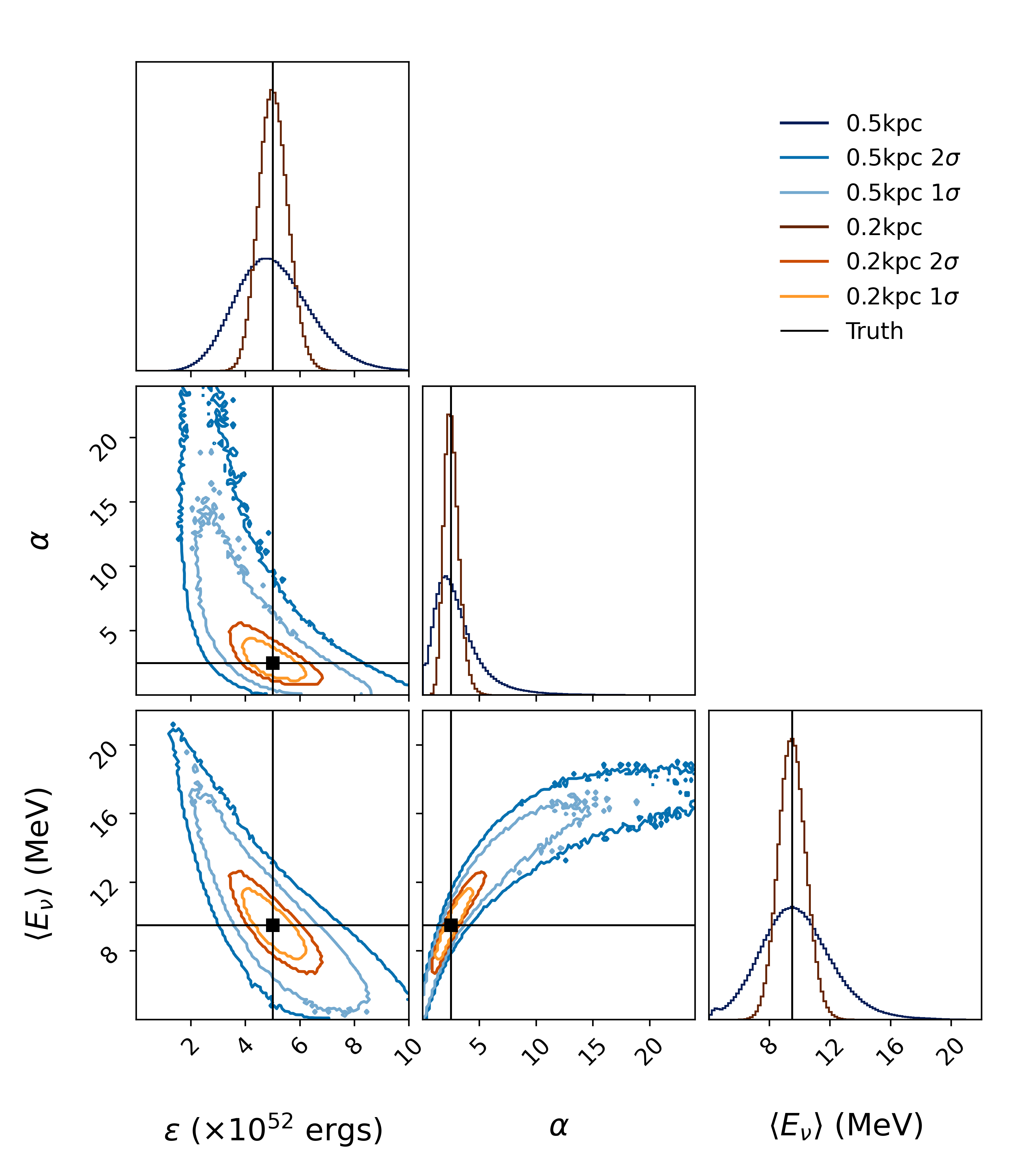}
	\caption{One- and two-sigma contours are drawn for the posterior probability distributions for CCSNe originating 0.2 and 0.5\,kpc from Earth, utilizing $1 \times 10^6$ datasets generated with Poisson statistics. The nominal set of parameters $(\alpha,\varepsilon,\langle E_{\nu}\rangle)$ of (2.5, 5$\times 10^{52}$\,erg, 9.5\,MeV) are indicated by the black lines.}
	\label{Fig:nexoCorner}
\end{figure}

Following each fit to a test dataset, the difference between the minimum negative log-likelihood ($LL_{\mbox{min}}$) and the negative log-likelihood calculated using the nominal set of parameters ($LL_{\mbox{true}}$) was determined, referred to as $-\Delta LL = -(LL_{\mbox{min}} - LL_{\mbox{true}})$. Because of the observed non-Gaussian nature of the distribution of $-\Delta LL$ values, one- and two-$\sigma$ confidence intervals (C.I.s) were generated by profiling over the nuisance parameters and selecting $-\Delta LL$ values that encapsulate 68.2\% and 95.5\% of the datasets, as shown in Fig.\ref{Fig:nexoCorner}. These bands are plotted for datasets generated at 0.2 and 0.5,kpc using the \texttt{corner}\cite{corner} package. Black lines indicate the nominal parameters of the incident supernova neutrino spectrum. The 1D histograms show the posterior probability distributions for each parameter.

\begin{figure}[ht!]
	\centering
	\includegraphics[width=\linewidth]{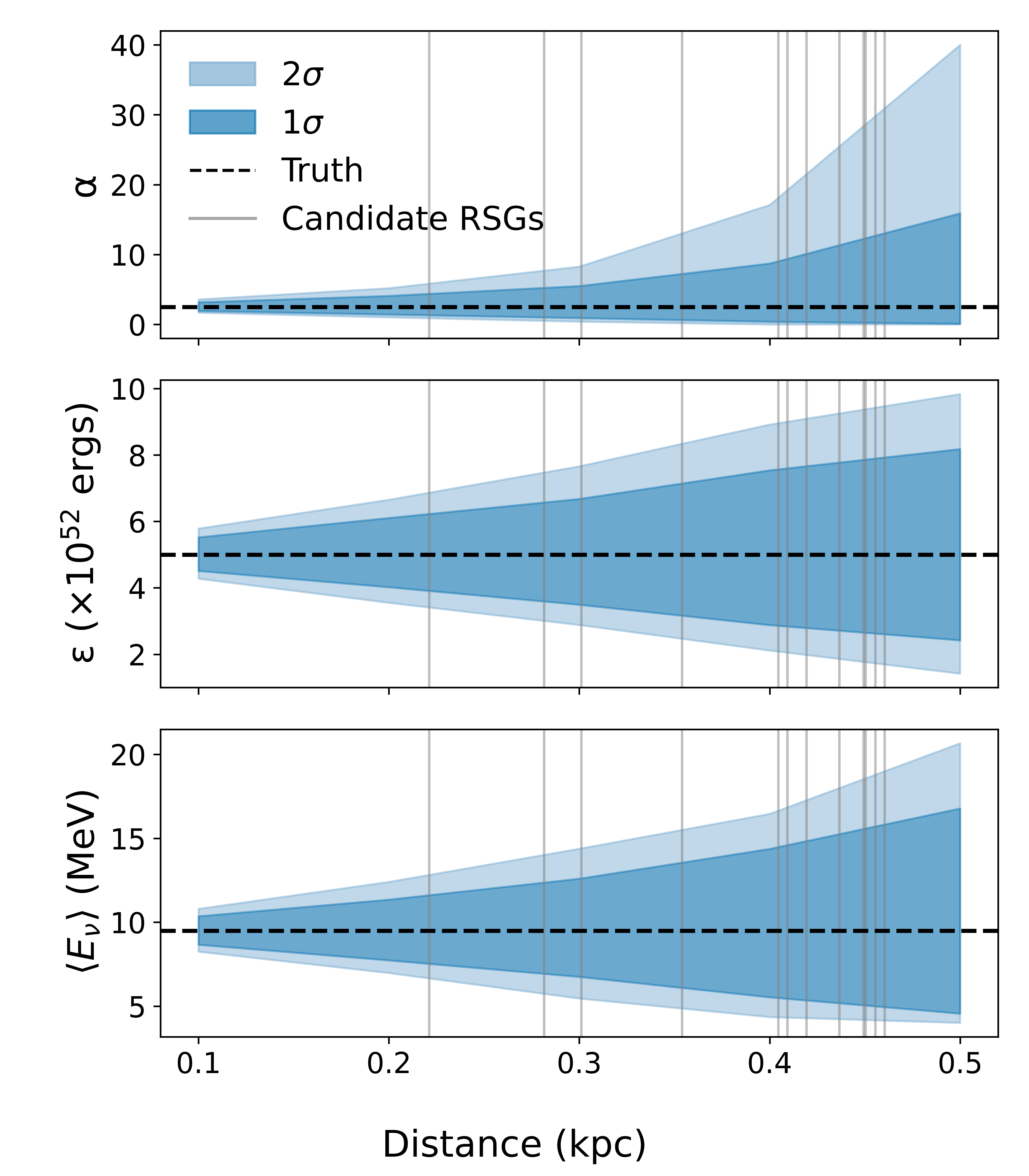}
	\caption{One- and two-$\sigma$ confidence intervals showing nEXO's ability to reconstruct incident electron-neutrino parameters as a function of distance. The dashed black line indicates the true parameter values, and the vertical gray lines indicate the location of nearby CCSNe candidates, from Ref.~\cite{Healy2024}.}
	\label{Fig:nexoDistance}
\end{figure}

To study nEXO's ability to reconstruct incident electron neutrino parameters as a function of CCSN distance, 60,000 datasets are generated with a step size of 0.1\,kpc. One- and two-$\sigma$ confidence intervals are formed by profiling over the nuisance parameters and selecting $-\Delta LL$ values enclosing 68.2\% and 95.5\% of the data, which are plotted in Fig.~\ref{Fig:nexoDistance} along with the true values of these parameters. The gray vertical lines show the locations of nearby RSGs, from Ref.~\cite{Healy2024}. At greater distances, many of the datasets do not have a sufficient number of events to capture the high-energy tail of the pinched-thermal distribution, resulting in a large increase in the spread of the $\alpha$ parameter. While limited statistics impact nEXO's ability to reconstruct the incident neutrino spectrum parameters, data from nEXO could be combined with that from other detectors to further constrain these parameters when statistics are poor.

\begin{figure}[ht]
	\centering
	\includegraphics[width=\linewidth]{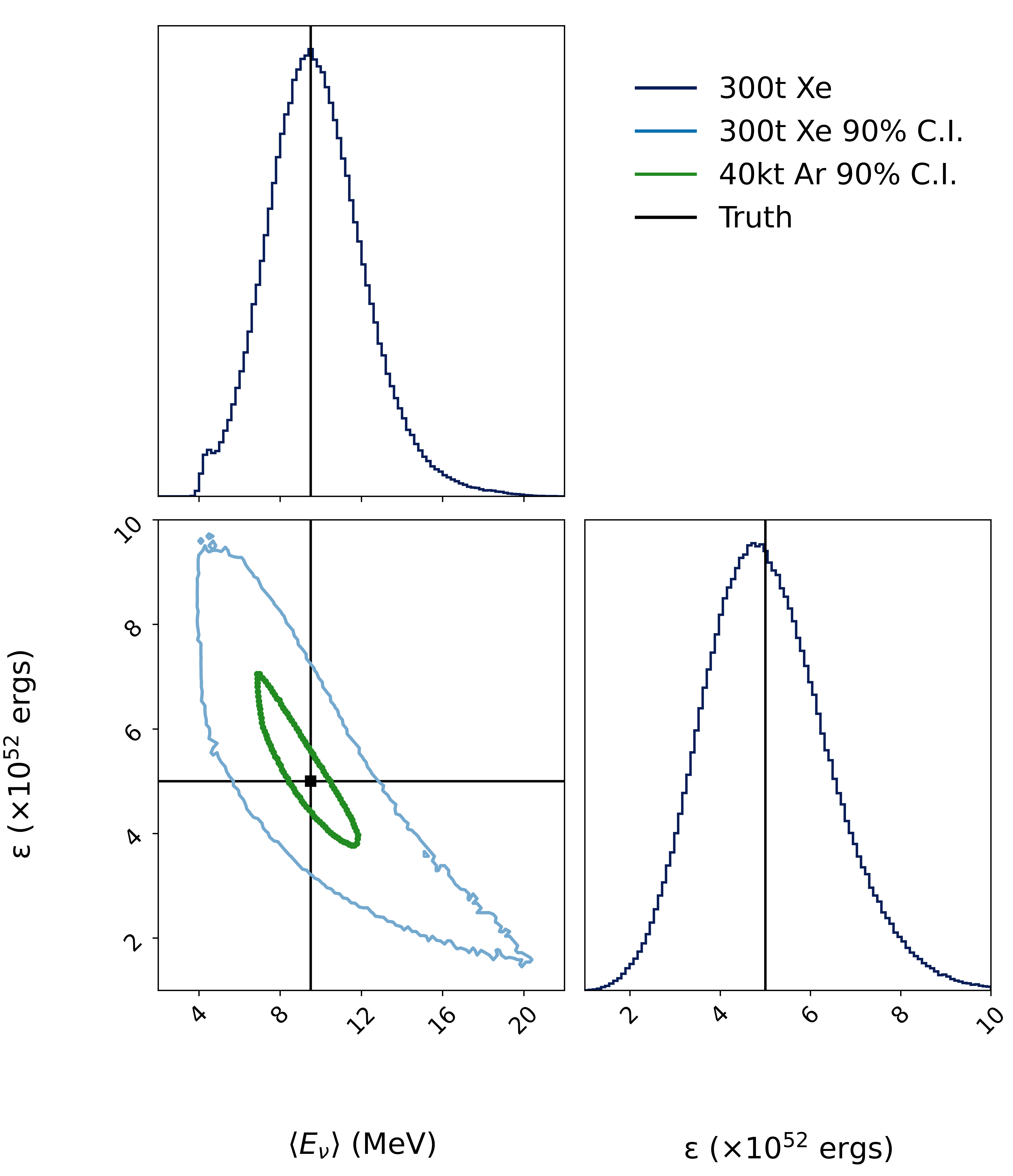}
	\caption{Distribution of posterior probability distributions for parameters are shown in the one-dimensional histograms, along with median values of the distribution, for a 300-metric-ton liquid xenon detector (90\% ${}^{136}$Xe, 10\% ${}^{134}$Xe) reconstructing events from a supernova located 4\,kpc away. The two-dimensional plot shows 90\% confidence intervals for this detector, as well as for a 40-metric-kt liquid argon detector, from Ref.~\cite{Abi2021}.}
	\label{Fig:300t}
\end{figure}

These results are extended to a 300-metric-ton liquid xenon detector based on the same isotopic abundance using the same transfer matrix. While this is an approximation, as the ratio of xenon skin to TPC volume will likely be different for a 300-metric-ton detector, Fig.~\ref{Fig:300t} illustrates how a detector of this size would compare to the 40-metric-kt liquid argon detector considered in Ref.~\cite{Abi2021}. While there are a number of challenges that need to be addressed to develop a xenon detector of this size~\cite{Avasthi2021}, with its increased mass compared to that of nEXO, a 300-metric-ton detector can reconstruct the incident neutrino spectrum to much further distances, and reconstruct nearby supernova electron-neutrino parameters with improved accuracy.

\subsection{Future modeling improvements}
The findings of this study can potentially be impacted by a number of factors whose contributions are difficult to quantify. The cross section and its energy dependence can be impacted by the value of the nuclear matrix elements and contributions from forbidden transitions not accounted for in the MARLEY model. While the calculation using MARLEY shows a general agreement with the calculations in Ref.~\cite{Pirinen2019}, uncertainties could be better quantified if experimental data existed for electron neutrinos in a similar energy range. The electron-neutrino spectrum shape and normalization impact the expected number of events that would be observed in nEXO. While only a handful of supernova neutrino events have been previously observed~\cite{Hirata1987,Bionta1987,Aglietta1987}, future supernova neutrino interactions would test these models. Although uncertainties on the spins, parities, and branching ratios of excited states of ${}^{136}$Cs are not expected to have a large impact on the visible spectrum from supernova neutrino interactions, experimental measurements of these quantities would lead to higher fidelity simulations. Finally, there may be additional energy deposited from electron-antineutrino interactions and NC interactions that can affect the ability to reconstruct the true electron-neutrino parameters, although these contributions are also expected to be small~\cite{Pirinen2018,Pirinen2019}.

%%%%%%%%%%%%%%%%%%%%%%%%%%%%%%%%%%%%%%%%%%%%%%%%%%%%%%%%%%%%%%%%%%%%%%
\section{Conclusion}
There is a wealth of physics that can be unlocked by studying the neutrinos from the next nearby core-collapse supernova, provided there are detectors capable of observing these interactions whose response is well known. We have evaluated the sensitivity of the nEXO detector to electron-neutrino interactions from supernovae using the MARLEY event generator, simulating CC interactions on the ${}^{134}$Xe and ${}^{136}$Xe isotopes. Events generated by MARLEY have been simulated using the nEXO simulation framework, and reconstructed using a custom algorithm designed to accommodate the unique nature of CC events compared to the signal and background events of primary interest to nEXO. Based on our work, we estimate that nEXO will be able to observe electron-neutrino interactions with xenon from supernovae as far as 5-8\,kpc from Earth, depending on the supernova neutrino flux model used and parameters of the collapsing star.

A transfer matrix was developed to map incident electron-neutrino energy into observed energy in nEXO, which is used to reconstruct the incident supernova neutrino spectrum using a pinched-thermal parameterization. We find that uncertainties in the parameter reconstruction grow rapidly for distances above $\sim$0.5kpc, where insufficient statistics in the tail of the spectrum limit the ability to reconstruct the pinching parameter. 

These results were extended to a 300-metric-ton xenon detector of the same isotopic abundance, which is able to detect supernova electron neutrinos out to further distances, covering most of the candidates listed within a recent survey of RSGs. While a 300-metric-ton detector would not be able to reconstruct supernova neutrino parameters with as much precision as DUNE, it would still provide useful information about the electron-neutrino component of nearby CCSNe and contribute to the global dataset of all supernova neutrino detectors.

\section{Acknowledgments}
 We would like to thank the authors of Ref.~\cite{Moreno2006} for providing their calculations of the GT strength used in this publication, and Steven Gardiner for his continued development and support of the MARLEY event generator. The authors gratefully acknowledge support for nEXO from the Office of Nuclear Physics within DOE's Office of Science, and NSF in the U.S.; from NSERC, CFI, FRQNT, NRC, and the McDonald Institute (CFREF) in Canada; from IBS in Korea; and from CAS and NSFC in China. This work was supported in part by Laboratory Directed Research and Development (LDRD) programs at Brookhaven National Laboratory (BNL), Lawrence Livermore National Laboratory (LLNL), Oak Ridge National Laboratory (ORNL), Pacific Northwest National Laboratory (PNNL), and SLAC National Accelerator Laboratory. This work was performed under the auspices of the U.S. Department of Energy by Lawrence Livermore National Laboratory under Contract No. DE-AC52-07NA27344.

\bibliography{main}% Produces the bibliography via BibTeX.

\clearpage\newpage
\appendix

%%%%%%%%%%%%%%%%%%%%%%%%%%
\section{${}^{132}\mbox{Xe}$ inclusive cross section comparison}\label{appendixC}
\begin{figure}[ht!]
	\centering
	\includegraphics[width=0.99\linewidth]{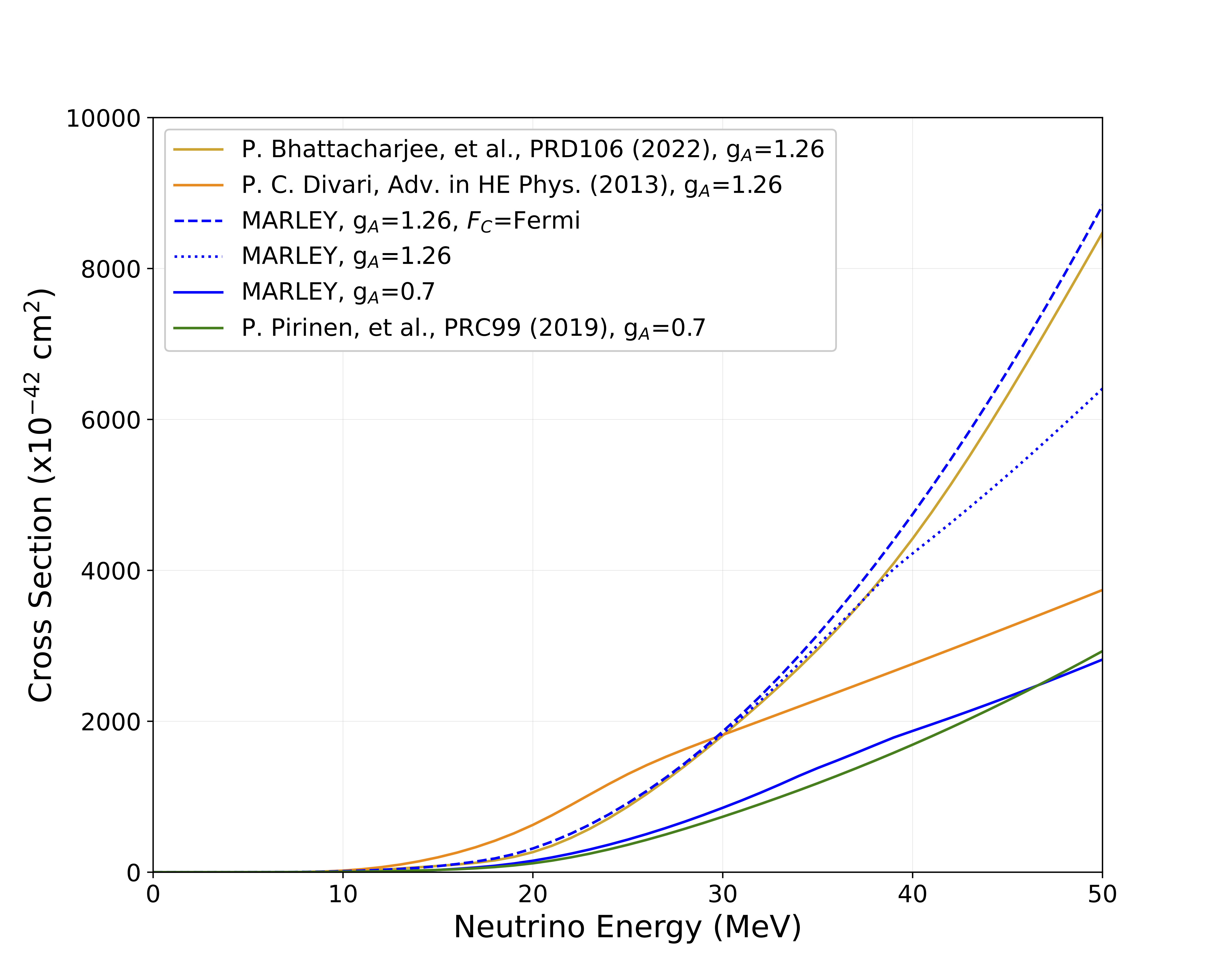}
	\caption{Predictions for the \XeTCC{} cross section from MARLEY using GT matrix elements from Ref.~\cite{Moreno2006} are compared with predictions from Refs.~\cite{Bhattacharjee2022,Pirinen2019,Divari2013} using different quenched $g_{A,eff}$ values and different Coulomb correction functions.}
	\label{fig:xe132}
\end{figure}

As a further test of MARLEY, we calculate predictions for the \XeTCC{} cross section as a function of energy using GT strength from Ref.~\cite{Moreno2006} and an IAS location predicted by Eq.~\ref{Eq:IAS}. Figure~\ref{fig:xe132} compares MARLEY's predictions with existing calculations for ${}^{132}$Xe from Refs.~\cite{Divari2013,Bhattacharjee2022,Pirinen2019}. References~\cite{Divari2013,Pirinen2019} use independent theoretical calculations that include forbidden transitions and use an approach to modeling the Coulomb correction similar to that used by MARLEY. Reference~\cite{Divari2013} assumes an unquenched $g_A$ of $\sim$1.26. Reference ~\cite{Bhattacharjee2022} uses the same theoretical GT matrix elements as we use in MARLEY, but assumes a different location for the IAS in ${}^{132}$Cs, an unquenched $g_A$ of $\sim$1.26, and uses a different approach to modeling the Coulomb correction. To demonstrate the impact of the Coulomb correction function and $g_A$ quenching, we modify these in MARLEY for the predictions in Fig~\ref{fig:xe132}. Details on MARLEY's approach to the Coulomb correction function can be found in Ref.~\cite{Gardiner2021}.

%%%%%%%%%%%%%%%%%%%%%%%%%%
\section{Individual output channel energy distributions}\label{appendixA}
\begin{figure}[ht!]
	\centering
	\includegraphics[width=0.99\linewidth]{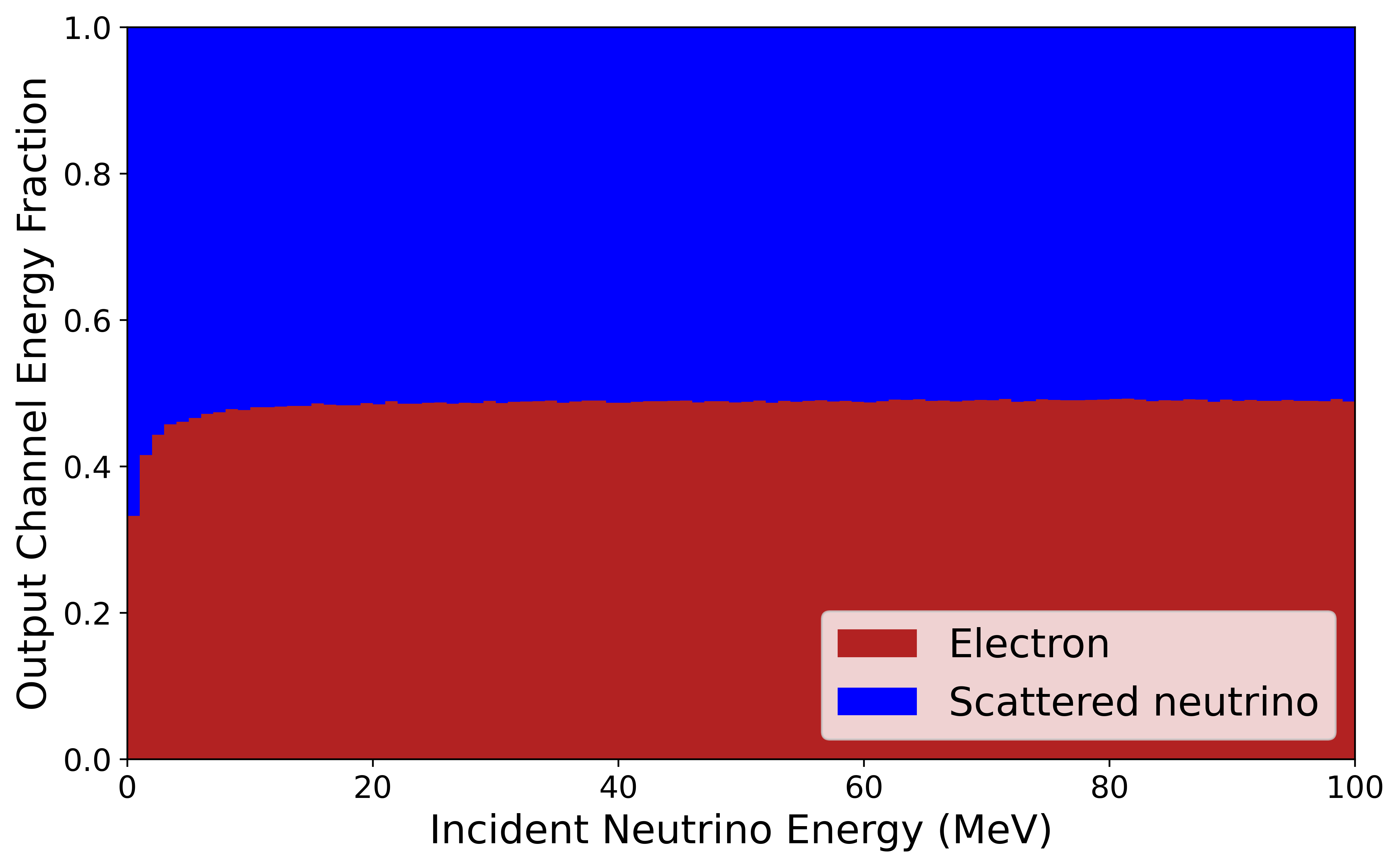}
	\caption{Output channel energy fraction for $\nu_e-e^-$.}
	\label{fig:esDist}
\end{figure}

\begin{figure}[ht!]
	\centering
	\includegraphics[width=0.99\linewidth]{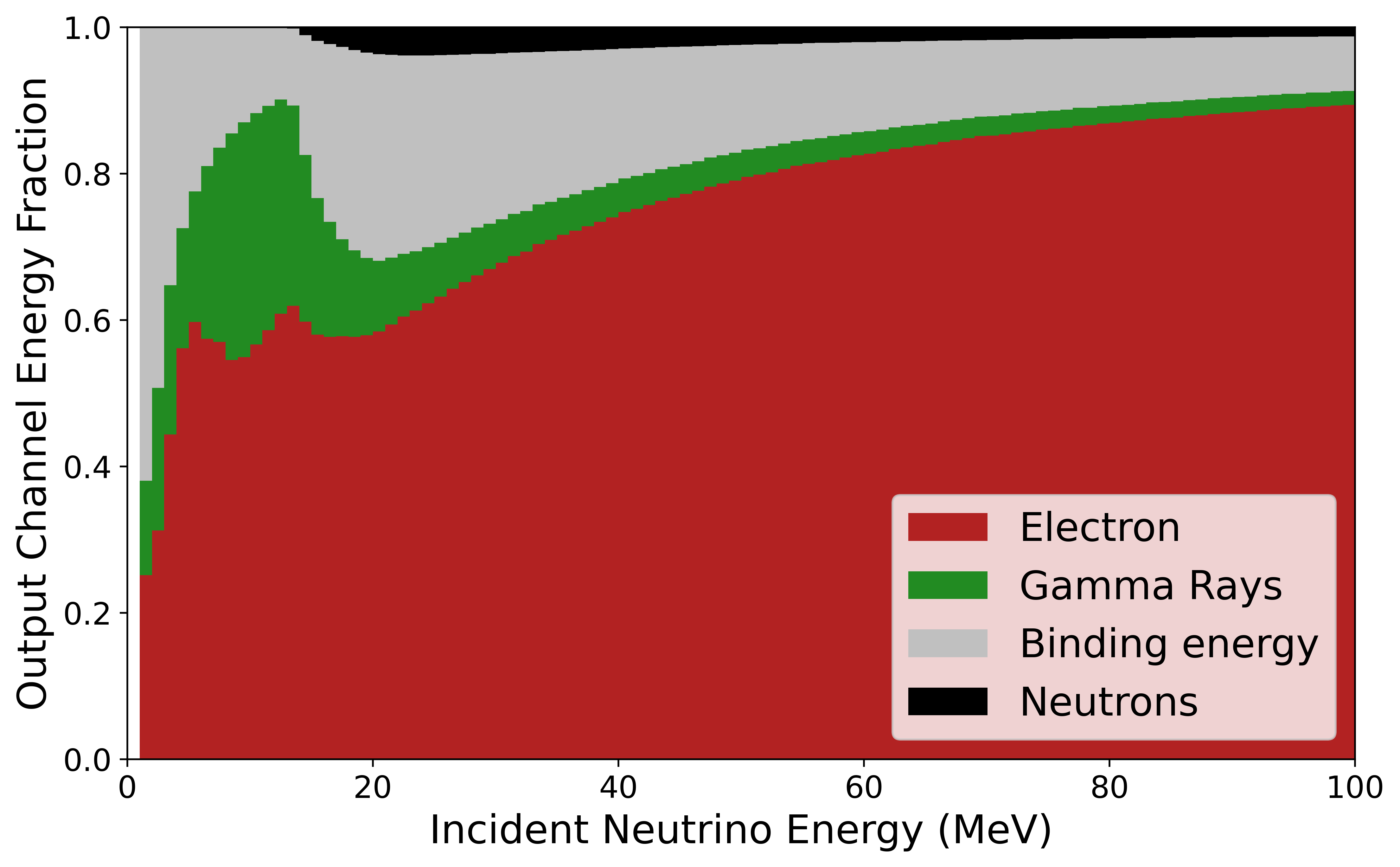}
	\caption{Output channel energy fraction for \XeFCC{}.}
	\label{fig:xe134Dist}
\end{figure}
\begin{figure}[ht!]
	\centering
	\includegraphics[width=0.99\linewidth]{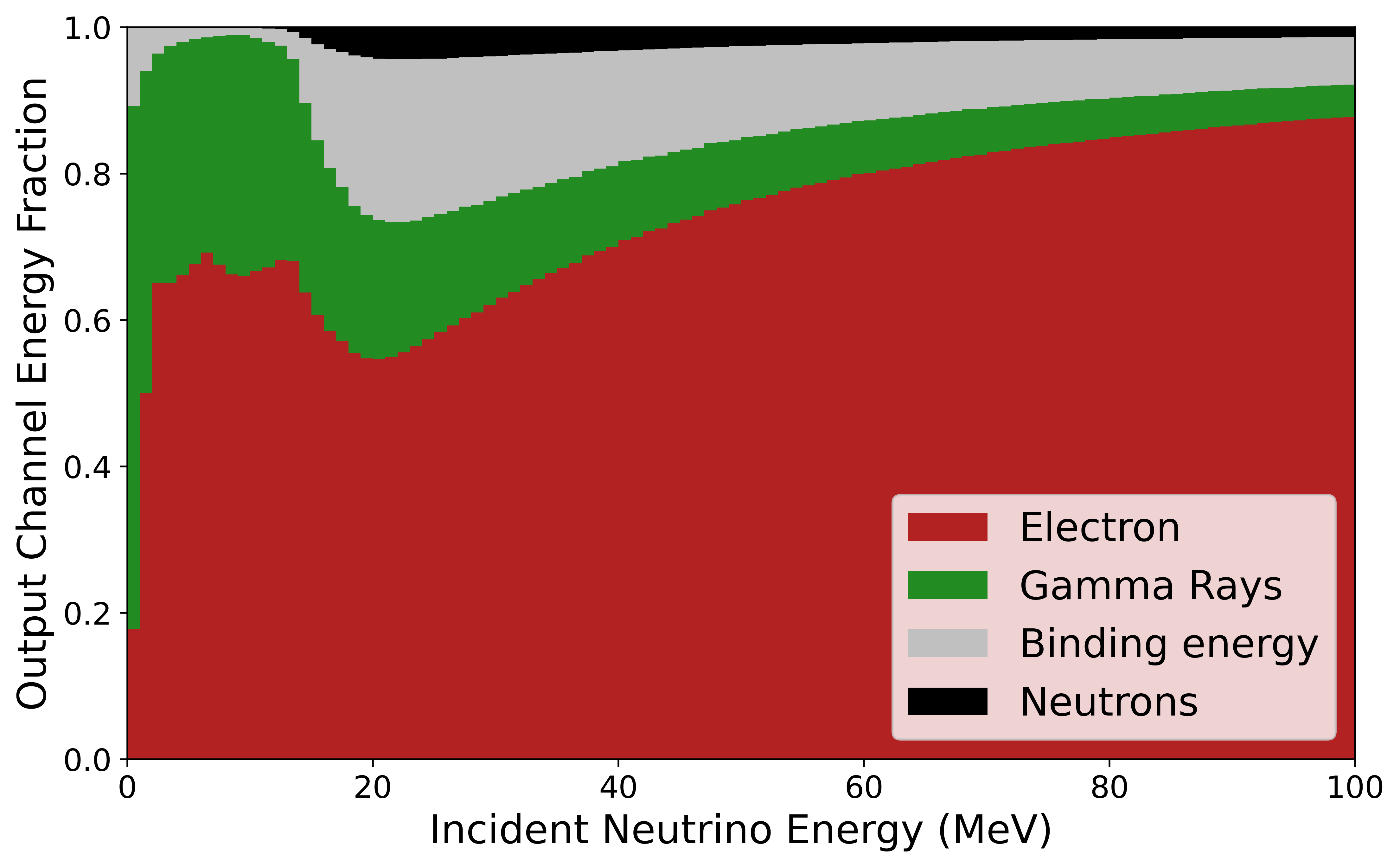}
	\caption{Output channel energy fraction for \XeSCC{}.}
	\label{fig:xe136Dist}
\end{figure}

Figs. ~\ref{fig:esDist}, \ref{fig:xe134Dist}, and \ref{fig:xe136Dist} show the fraction of the incident neutrino energy transferred to the various output channels for \veES{}, \XeFCC{}, and \XeSCC{} events. 

%%%%%%%%%%%%%%%%%%%%%%%%%%

\section{Individual transfer matrices}\label{appendixB}

\begin{figure}[ht!]
	\centering
	\includegraphics[width=0.95\linewidth]{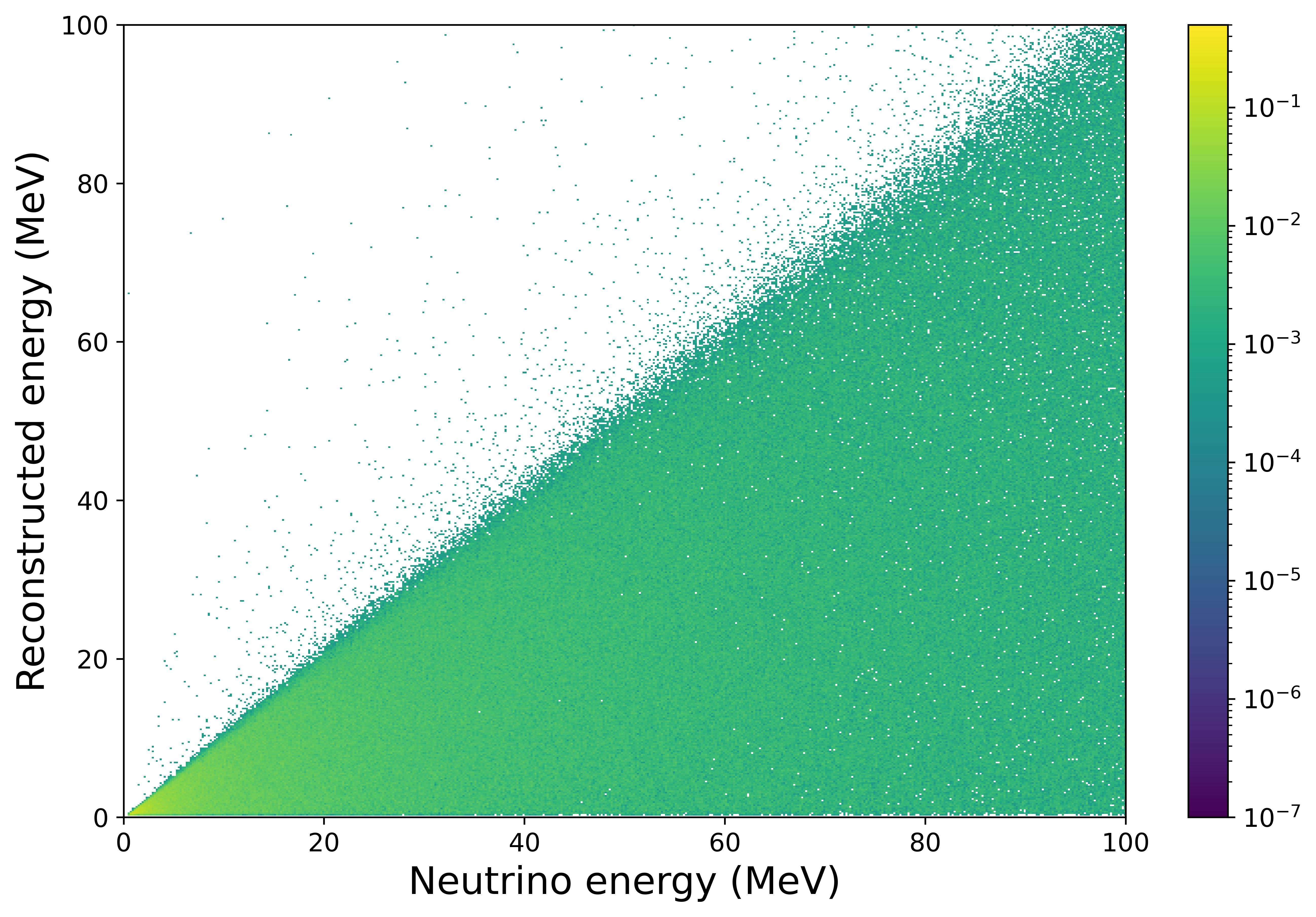}
	\caption{Transfer matrix for $\nu_e-e^-$.}
	\label{fig:esTransfer}
\end{figure}

\begin{figure}[ht!]
	\centering
	\includegraphics[width=0.95\linewidth]{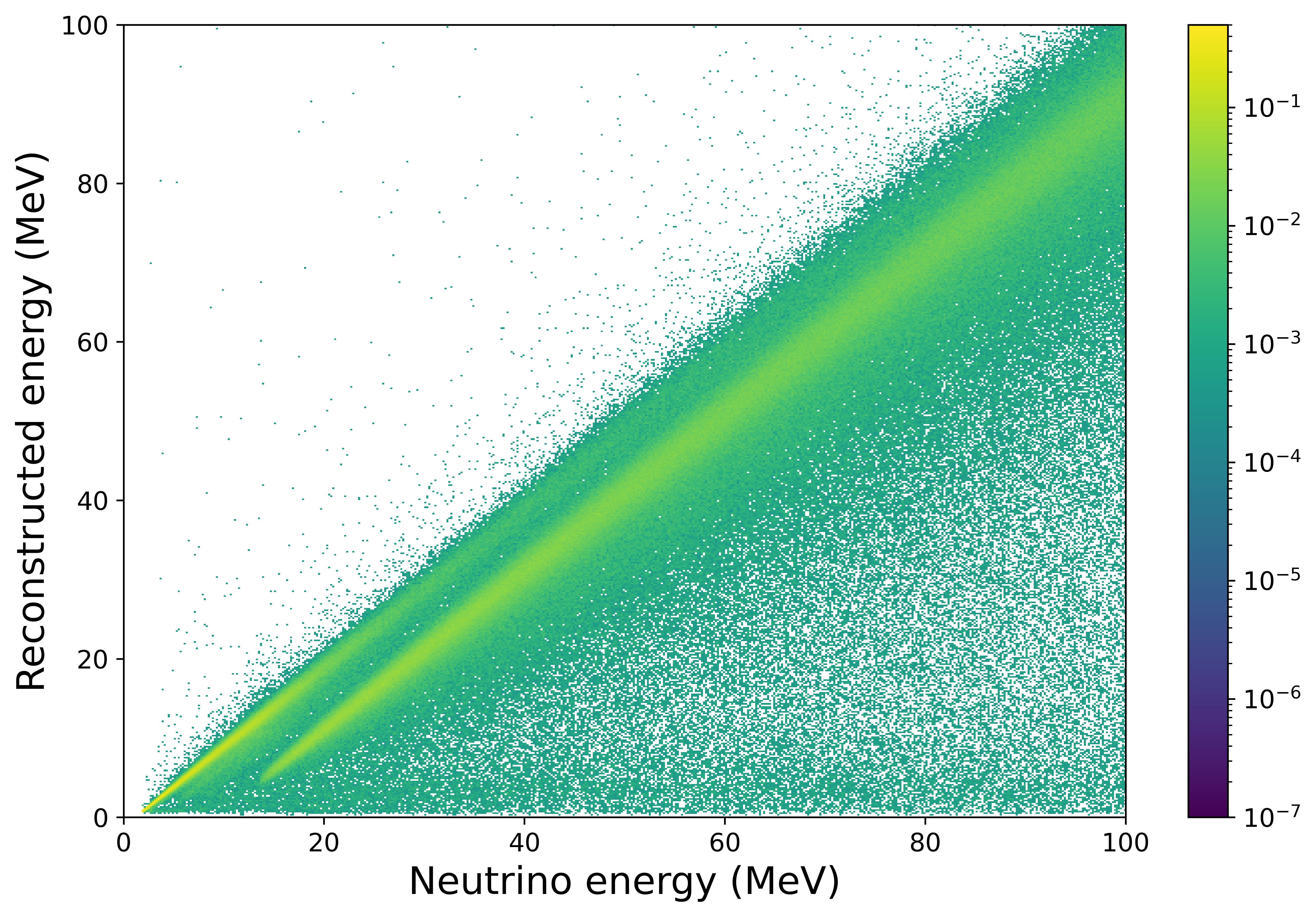}
	\caption{Transfer matrix for \XeFCC{}.}
	\label{fig:xe134Transfer}
\end{figure}

\begin{figure}[ht!]
	\centering
	\includegraphics[width=0.95\linewidth]{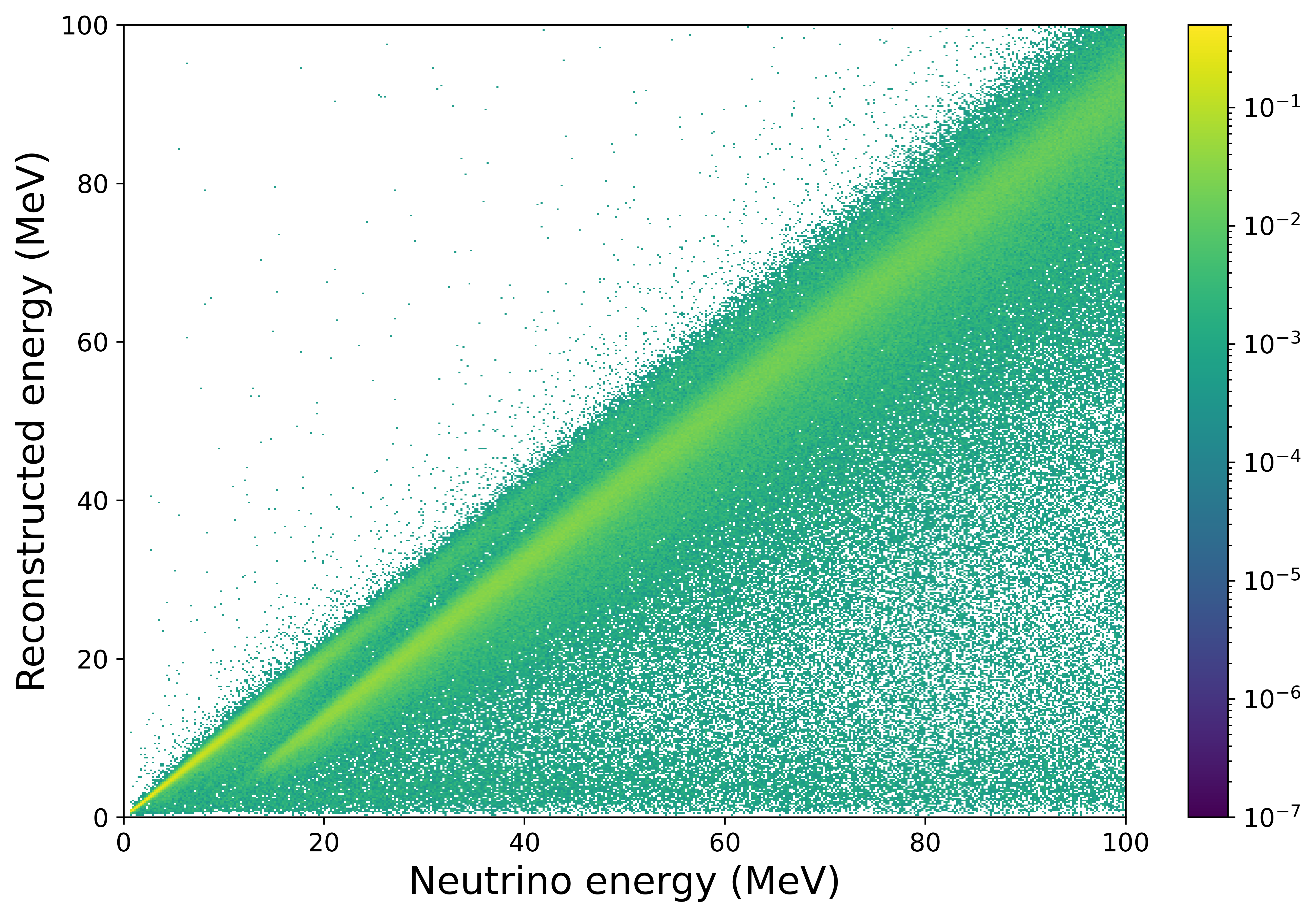}
	\caption{Transfer matrix for \XeSCC{}.}
	\label{fig:xe136Transfer}
\end{figure}

Figs.~\ref{fig:esTransfer},\ref{fig:xe134Transfer}, and\ref{fig:xe136Transfer} show the transfer matrices for each of the scattering components described within the text, generated with $1 \times 10^6$ simulated and reconstructed events with a uniform neutrino distribution from 0 to 100\,MeV. These are combined with isotopic and cross section weighting in Fig.~\ref{Fig:combined_transferMatrix} of the main text.

\end{document}

%% file: authors.tex
\newcommand{\UCSD}{\affiliation{Physics Department, University of California San Diego, La Jolla, CA 92093, USA}}
\newcommand{\McGill}{\affiliation{Physics Department, McGill University, Montr\'eal, QC H3A 2T8, Canada}}
\newcommand{\Stanford}{\affiliation{Physics Department, Stanford University, Stanford, CA 94305, USA}}
\newcommand{\Erlangen}{\affiliation{Erlangen Centre for Astroparticle Physics (ECAP), Friedrich-Alexander University Erlangen-N{\"u}rnberg, Erlangen 91058, Germany}}
\newcommand{\PNNL}{\affiliation{Pacific Northwest National Laboratory, Richland, WA 99352, USA}}
\newcommand{\Carleton}{\affiliation{Department of Physics, Carleton University, Ottawa, ON K1S 5B6, Canada}}
\newcommand{\UMass}{\affiliation{Amherst Center for Fundamental Interactions and Physics Department, University of Massachusetts, Amherst, MA 01003, USA}}
\newcommand{\ITEP}{\affiliation{National Research Center ``Kurchatov Institute'', Moscow, 123182, Russia}}
\newcommand{\LLNL}{\affiliation{Lawrence Livermore National Laboratory, Livermore, CA 94550, USA}}
\newcommand{\UK}{\affiliation{Department of Physics and Astronomy, University of Kentucky, Lexington, KY 40506, USA}}
\newcommand{\BNL}{\affiliation{Brookhaven National Laboratory, Upton, NY 11973, USA}}
\newcommand{\SLAC}{\affiliation{SLAC National Accelerator Laboratory, Menlo Park, CA 94025, USA}}
\newcommand{\RPI}{\affiliation{Department of Physics, Applied Physics, and Astronomy, Rensselaer Polytechnic Institute, Troy, NY 12180, USA}}
\newcommand{\Laurentian}{\affiliation{School of Natural Sciences, Laurentian University, Sudbury, ON P3E 2C6, Canada}}
\newcommand{\IHEP}{\affiliation{Institute of High Energy Physics, Chinese Academy of Sciences, Beijing, 100049, China}}
\newcommand{\IME}{\affiliation{Institute of Microelectronics, Chinese Academy of Sciences, Beijing, 100029, China}}
\newcommand{\Sherbrooke}{\affiliation{Universit\'e de Sherbrooke, Sherbrooke, QC J1K 2R1, Canada}}
\newcommand{\Alabama}{\affiliation{Department of Physics and Astronomy, University of Alabama, Tuscaloosa, AL 35405, USA}}
\newcommand{\UNCW}{\affiliation{Department of Physics and Physical Oceanography, University of North Carolina Wilmington, Wilmington, NC 28403, USA}}
\newcommand{\UBC}{\affiliation{Department of Physics and Astronomy, University of British Columbia, Vancouver, BC V6T 1Z1, Canada}}
\newcommand{\TRIUMF}{\affiliation{TRIUMF, Vancouver, BC V6T 2A3, Canada}}
\newcommand{\Drexel}{\affiliation{Department of Physics, Drexel University, Philadelphia, PA 19104, USA}}
\newcommand{\ORNL}{\affiliation{Oak Ridge National Laboratory, Oak Ridge, TN 37831, USA}}
\newcommand{\CSU}{\affiliation{Physics Department, Colorado State University, Fort Collins, CO 80523, USA}}
\newcommand{\Yale}{\affiliation{Wright Laboratory, Department of Physics, Yale University, New Haven, CT 06511, USA}}
\newcommand{\USD}{\affiliation{Department of Physics, University of South Dakota, Vermillion, SD 57069, USA}}
\newcommand{\Mines}{\affiliation{Department of Physics, Colorado School of Mines, Golden, CO 80401, USA}}
\newcommand{\CUP}{\affiliation{IBS Center for Underground Physics, Daejeon, 34126, Korea}}
\newcommand{\UWC}{\affiliation{Department of Physics and Astronomy, University of the Western Cape, P/B X17 Bellville 7535, South Africa}}
\newcommand{\SUBATECH}{\affiliation{SUBATECH, Nantes Universit\'e, IMT Atlantique, CNRS/IN2P3, Nantes 44307, France}}
\newcommand{\Caltech}{\affiliation{California Institute of Technology, Pasadena, CA 91125, USA}}
\newcommand{\Bern}{\affiliation{LHEP, Albert Einstein Center, University of Bern, 3012 Bern, Switzerland}}
\newcommand{\Skyline}{\affiliation{Skyline College, San Bruno, CA 94066, USA}}
\newcommand{\SNOLAB}{\affiliation{SNOLAB, Lively, ON P3Y 1N2, Canada}}
\newcommand{\Muenster}{\affiliation{Institut f{\"u}r Kernphysik, Westf{\"a}lische Wilhelms-Universit{\"a}t M{\"u}nster, M{\"u}nster 48149, Germany}}
\newcommand{\FRIB}{\affiliation{Facility for Rare Isotope Beams, Michigan State University, East Lansing, MI 48824, USA}}
\newcommand{\Queens}{\affiliation{Department of Physics, Queen's University, Kingston, ON K7L 3N6, Canada}}
\newcommand{\Windsor}{\affiliation{Department of Physics, University of Windsor, Windsor, ON N9B 3P4, Canada}}
\newcommand{\TUM}{\affiliation{Physikdepartment and Excellence Cluster Universe, Technische Universit{\"a}t M{\"u}nchen, Garching 80805, Germany}}

\author{S.~Hedges}\email{hedges3@llnl.gov}\LLNL
\author{S.~Al~Kharusi}\Stanford\McGill
\author{E.~Angelico}\Stanford
\author{J.~P.~Brodsky}\LLNL
\author{G.~Richardson}\Yale
\author{S.~Wilde}\Yale

\author{A.~Amy}\SUBATECH
\author{A.~Anker}\SLAC
\author{I.~J.~Arnquist}\PNNL
\author{P.~Arsenault}\Sherbrooke
% ...
